\documentclass[acmsmall,screen,nonacm]{acmart}
\AtBeginDocument{
  }

\usepackage[dvipsnames]{xcolor}

\usepackage[group-separator={,}]{siunitx}
\usepackage{multirow}
\usepackage{array}
\usepackage{longtable}
\usepackage[table]{xcolor}
\usepackage{underscore} 
\usepackage{url}
\usepackage{hyperref}
\usepackage{subfig}
\usepackage{graphicx}
\usepackage{subcaption}
\usepackage{rotating}

\usepackage[edges]{forest}
\usepackage{xcolor}

\usepackage[utf8]{inputenc}
\usepackage{forest}
\usepackage{tikz}

\usetikzlibrary{shadows}

\usepackage{listings}
\usepackage[most]{tcolorbox}
\definecolor{bg}{RGB}{248,248,248}
\definecolor{frame}{RGB}{210,210,210}
\definecolor{kw}{HTML}{1750EB}
\definecolor{str}{HTML}{067D17}
\definecolor{ph}{RGB}{175,95,0}
\definecolor{mainfindingsbg}{RGB}{240,241,248}
\definecolor{mainfindingsframe}{RGB}{0,0,104}
\definecolor{keyfindingsbg}{RGB}{239,248,255}
\definecolor{keyfindingsframe}{RGB}{93,156,205}
\newcolumntype{L}[1]{>{\raggedright\arraybackslash}p{#1}}
\newcommand{\kw}[1]{\textcolor{kw}{#1}}

\newtcolorbox{mainfindingsbox}[1]{
    enhanced,
    unbreakable,
    colback=mainfindingsbg,
    colframe=mainfindingsframe,
    coltitle=white,
    fonttitle=\bfseries,
    title=#1,
    boxrule=0.8pt,
    arc=1mm,
    left=1.5mm,
    right=1.5mm,
    top=1.2mm,
    bottom=1.2mm,
    before skip=0.9em,
    after skip=0.9em,
    before upper={
        \setlength{\leftmargini}{1.15em}
        \setlength{\itemsep}{0.2em}
        \setlength{\parsep}{0pt}
        \setlength{\topsep}{0.2em}
    }
}
\newtcolorbox{keyfindingsbox}[1]{
    enhanced,
    unbreakable,
    colback=keyfindingsbg,
    colframe=keyfindingsframe,
    coltitle=white,
    fonttitle=\bfseries,
    title=#1,
    boxrule=0.5pt,
    arc=1mm,
    left=1mm,
    right=1mm,
    top=1mm,
    bottom=1mm,
    before skip=0.8em,
    after skip=0.8em,
    before upper={
        \setlength{\leftmargini}{1.15em}
        \setlength{\itemsep}{0.15em}
        \setlength{\parsep}{0pt}
        \setlength{\topsep}{0.15em}
    }
}
\begin{document}

\title{Real Faults in Model Context Protocol (MCP) Software: a Comprehensive Taxonomy}

\author{Mina Taraghi}
\authornote{Both authors contributed equally to this research.}
\email{mina.taraghi@polymtl.ca}
\orcid{0009-0007-8250-4200}
\affiliation{
  \institution{Polytechnique Montreal}
  \city{Montreal}
  \country{Canada}
}

\author{Mohammad Mehdi Morovati}
\email{mehdi.morovati@polymtl.ca}
\authornotemark[1]
\affiliation{
  \institution{Polytechnique Montreal}
  \city{Montreal}
  \country{Canada}
}

\author{Foutse Khomh}
\email{foutse.khomh@polymtl.ca}
\affiliation{
  \institution{Polytechnique Montreal}
  \city{Montreal}
  \country{Canada}
}

\renewcommand{\shortauthors}{Taraghi et al.}

\begin{abstract}

The rapid adoption of foundation models has significantly expanded the capabilities of software systems, enabling them to perform complex language, reasoning, and interaction tasks that were previously difficult to automate. However, this progress has also introduced novel challenges that were largely absent in previous generations of software. In particular, the increasing integration of foundation models with external tools and resources raises new concerns regarding reliability, security, and robustness. The Model Context Protocol (MCP) has recently been proposed to standardize interactions between AI-based software systems, software tools, and external resources. Despite its growing adoption, there remains limited systematic understanding of real-world faults in MCP-based software systems.

In this paper, we present the first large-scale taxonomy of faults in MCP servers, comprising five high-level fault categories derived from empirical evidence. To evaluate the completeness and generalizability of this taxonomy, we conduct a survey of MCP practitioners representing diverse professional roles. The results confirm that all MCP-specific fault categories occur in practice and reveal distinct characteristics that differentiate MCP-specific faults from non-MCP faults.

Overall, these findings provide actionable guidance for building more robust MCP-based systems through boundary-aware testing, response validation, configuration diagnostics, and reproducible issue reporting.

\end{abstract}



\maketitle

\section{Introduction}
Foundation models, and more specifically Large Language Models (LLMs), represent a transformative advancement in Artificial Intelligence (AI), demonstrating near human-level performance across a wide range of tasks, such as reasoning and decision-making~\cite{naveed2025comprehensive}. This remarkable capability has motivated developers and researchers to integrate LLMs into diverse domains, including medical science~\cite{shool2025systematic}, coding~\cite{dong2025survey}, media~\cite{meier2024llm}, etc.
As a result, software systems that incorporate LLMs as core components to enhance functionality and performance are commonly referred to as LLM-based software systems.

The current integration of LLMs into software systems typically relies on customized Application Programming Interfaces (APIs) that allow LLMs to communicate with external tools and data sources. However, the lack of standardization in API design introduces several challenges, such as increased integration complexity, limited scalability across heterogeneous sources, and reduced interoperability~\cite{singh2025survey}. To mitigate these issues, the Model Context Protocol (MCP) has been introduced as an open protocol that standardizes the communication between applications and LLMs, thereby simplifying integration and promoting interoperability~\cite{mcp_introduction}.

In contrast to traditional software systems, whose outputs are produced through deterministic logic, LLM-based software systems rely on probabilistic reasoning processes that generate dynamic, context-dependent behaviors~\cite{ha2025evaluating}. Owing to this fundamental difference, engineering practices for LLM-based systems face distinct challenges and require specialized expertise.

As LLM-based software systems are increasingly deployed in safety-critical domains, such as financial services~\cite{yu2024fincon} and autonomous driving~\cite{li2024large}, ensuring their reliability and robustness becomes imperative. However, because of the intrinsic differences between traditional and LLM-based systems, existing testing techniques designed for traditional software are often insufficient for LLM-driven applications. Consequently, there is a growing need for novel testing methodologies specifically tailored to LLM-based software systems. A fundamental step toward this goal is understanding the characteristics of faults that arise in such systems, as identifying fault characteristics is central to testing any software system~\cite{morovati2024bug}. Therefore, investigating faults in LLM-based software systems is essential for developing effective testing techniques that are adapted to their unique properties. 

Although a few studies have examined faults in LLM-based systems more broadly~\cite{vinay_failure_2025,winston_taxonomy_2025,tambon2025bugs}, to the best of our knowledge, no prior research has specifically explored the characteristics of faults in MCP-based systems. To address this gap, this paper presents a large-scale empirical study that (1) introduces the first 
taxonomy of faults in MCP servers, (2) analyzes practitioners’ perspectives on MCP-related faults, and (3) characterizes differences among MCP fault categories as well as distinctions between MCP and non-MCP faults. Accordingly, we investigate the following research questions (RQ):
\begin{itemize}
\item[\textbf{RQ1.}] What types of faults commonly arise during the development of MCP servers?
\item[\textbf{RQ2.}] How do MCP server practitioners perceive the identified faults?
\item[\textbf{RQ3.}] What are the key characteristics of the various MCP-related faults?
\end{itemize}

To address these RQs, we first cluster 3,282 
bug-related GitHub closed issues extracted from repositories providing an MCP server, resulting in 407 MCP-related issues. We then manually label these issues to derive a taxonomy of faults observed in MCP server development. To validate the taxonomy, we conduct a survey with MCP practitioners. In addition, we analyze and report the distinctive characteristics of the identified MCP-related fault types. The main contributions of this study can be summarized as follows:

\begin{itemize}
    \item We conduct the first large-scale empirical study on real faults that arise during the development of MCP servers.
    \item We categorize MCP-related issues reported in MCP server repositories and derive a comprehensive taxonomy of these faults.
    \item We validate the proposed taxonomy through a survey of MCP practitioners.
    \item We highlight key differences among MCP-related faults by analyzing their distinctive characteristics.
    \item We translate the taxonomy, practitioner survey, and issue-characteristic analyses into actionable guidance for MCP developers, including boundary-aware testing, response-contract validation, transport and configuration checks, logging separation, and reproducible issue reporting.
\end{itemize}

\textbf{The rest of this paper is organized as follows}. 
Section~\ref{sec:background} presents the necessary background on MCP.
In Section~\ref{sec:method}, we describe the methodology employed in this study.
Section~\ref{sec:results} reports our empirical findings.
In Section~\ref{sec:relatedwork}, we discuss the most closely related work.
Finally, Section~\ref{sec:threat} outlines the threats to validity, and Section~\ref{sec:conclusion} concludes the paper.

\section{Background}
\label{sec:background}

\subsection{Large Language Model (LLM)}

In general, Artificial Intelligence (AI) refers to the study and development of systems that exhibit machine-based intelligence, as opposed to biological intelligence~\cite{mccarthy2007artificial}. A prominent subfield of AI is generative AI, which focuses on models capable of producing new content (e.g., text, images, or code). Within this subfield, Large Language Models (LLMs) constitute a family of foundation models trained on massive text corpora to generate and understand natural language. In recent years, LLMs have attracted significant attention from both academia and industry and have substantially advanced the state of the art in AI~\cite{minaee2024large,wei2022emergent}.
LLMs are now widely used as powerful tools for performing a diverse range of complex tasks, including Natural Language Processing (NLP), machine translation, computer vision, question answering, and more~\cite{hadi2023large}.
Owing to their broad capabilities, an increasing number of software systems across various domains (such as healthcare~\cite{thirunavukarasu2023large}, financial systems~\cite{li2023large}, and autonomous vehicles~\cite{fu2024drive}) are integrating LLMs to enhance their performance.

Although LLMs outperform many earlier AI approaches (e.g., Deep Learning (DL)), in several benchmarks and real-world tasks, they also introduce significant challenges~\cite{kaddour2023challenges}. First, their scale imposes substantial computational and infrastructural demands. The ever-growing size of pre-training datasets, the storage overhead of fine-tuned or adapted variants, and the large memory footprint required for loading and serving these models make their deployment and analysis impractical for many individual users and small organizations~\cite{liu2024datasets,li2021prefix}.

Second, LLMs are highly sensitive to input formulation. Even minor variations in prompt wording, structure, or formatting can lead to considerable differences in generated outputs, affecting both reliability and reproducibility~\cite{zhao2021calibrate,lu2021fantastically}.

\subsection{Model Context Protocol (MCP)}
The Model Context Protocol (MCP) is an open protocol that standardizes the interaction between applications and LLMs by formalizing how contextual information and external tool capabilities are provided to the model. MCP specifies a structured and model-agnostic interface through which external resources, such as file systems, databases, and web APIs, can securely expose their functionalities to LLM-based systems. This design enables the interoperable integration of LLMs within complex software environments~\cite{hou_model_2025,hasan2025model}.

In contrast to prior ad hoc integration mechanisms, including custom APIs, plugin-based extensions, or tightly coupled agent frameworks, MCP defines a unified abstraction that supports both information retrieval and action execution. It elevates essential mechanisms, such as capability discovery, schema negotiation, and access control, to first-class protocol constructs. By doing so, MCP promotes modularity, extensibility, and security in the engineering of LLM-enabled systems~\cite{hou_model_2025}.

MCP follows a host--client--server architecture composed of three core components~\cite{mcp_architecture,hou_model_2025}. The \textit{MCP host} is an AI application that orchestrates interactions with external systems while running one or more MCP clients; typical hosts include chat applications, IDEs, and autonomous agents~\cite{mcp_architecture}. The \textit{MCP client} operates within the host and maintains a one-to-one communication link with a corresponding MCP server, acting as an intermediary that discovers server capabilities, issues requests, and processes responses and notifications~\cite{mcp_clients}. The \textit{MCP server} is a standalone program, executing locally or remotely, that exposes external functionality and data to AI systems through standardized interfaces~\cite{mcp_servers}.

MCP servers provide three primary types of capabilities: \textit{tools}, \textit{resources}, and \textit{prompts}~\cite{mcp_tools,mcp_resources,mcp_prompts}. \textit{Tools} enable servers to perform external operations, such as invoking web APIs, accessing databases, or executing system commands, on behalf of the AI model~\cite{mcp_tools}. \textit{Resources} expose structured or unstructured data sources—including local files, databases, or cloud repositories—that can be retrieved as contextual input for model reasoning~\cite{mcp_resources}. \textit{Prompts} consist of reusable templates or workflows maintained by the server to standardize interactions and improve consistency across repeated tasks~\cite{mcp_prompts}.

From a protocol perspective, MCP is organized into two conceptual layers: a \textit{data layer} and a \textit{transport layer}~\cite{mcp_architecture}. The data layer defines the core JSON-RPC~2.0-based communication protocol~\cite{JSONRPCSpecification}, including message formats, lifecycle management, capability negotiation, and the semantics of MCP primitives such as tools, resources, prompts, and notifications~\cite{mcp_architecture}. The transport layer abstracts the underlying communication mechanisms and supports both local and remote interactions, including standard input/output (STDIO) for local processes and streamable HTTP for remote servers, along with authentication and authorization mechanisms~\cite{mcp_architecture}.

A typical MCP workflow begins with lifecycle initialization, during which the client and server negotiate protocol versions and supported capabilities~\cite{hou_model_2025,mcp_architecture}. Once connected, the client queries the server for its available tools, resources, and prompts, enabling the host to dynamically select and invoke appropriate capabilities based on user intent or task context~\cite{hou_model_2025}. Tool execution and data access follow a structured request--response pattern, while notifications allow servers to asynchronously report progress or state changes to the client, ensuring continuous synchronization between the AI application and external systems~\cite{hou_model_2025}.

\begin{figure}
    \centering
    \includegraphics[width=0.7\linewidth]{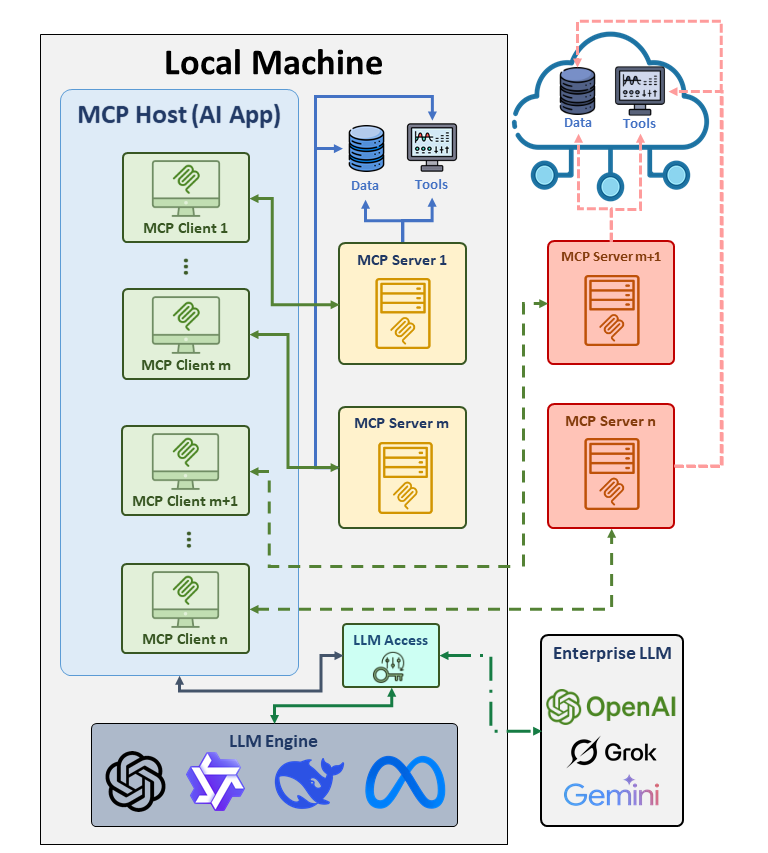}
    \caption{A high-level view of Model Context Protocol (MCP) Architecture
    }
    \label{fig:mcp-arch}
\end{figure}

To support adoption and development, MCP provides official software development kits (SDKs) for various programming languages \cite{mcp_sdks}, reference implementations, and developer tooling for building MCP clients and servers across different environments~\cite{mcp_servers,mcp_clients}. These SDKs abstract low-level protocol details, allowing developers to focus on implementing server capabilities or integrating MCP into host applications. While this ecosystem facilitates rapid development and interoperability, it also introduces new sources of complexity related to configuration, protocol compliance, and cross-component coordination, which motivates a systematic study of faults in MCP-based software.
Fig.\ref{fig:mcp-arch} represents a high-level view of MCP architecture.

\subsection{Faults in MCP servers}

A software fault is the manifestation of a software error, defined as an incorrect portion of the software introduced by a developer’s mistake that may lead to unexpected or incorrect system behavior~\cite{galin2004software}. In other words, a fault reflects a discrepancy between the intended requirements and the implemented functionality~\cite{ieee8016712}. When a user interacts with the faulty component, the fault may be activated, potentially causing a software failure~\cite{tian2005software}.

Like other software systems, MCP servers are subject to faults. However, their unique architecture and operational paradigm distinguish them from typical LLM-based applications~\cite{singh2025survey}, making them susceptible to specific categories of faults that are less common—or even absent—in other types of LLM-enabled systems.
In general, faults in MCP-based systems can originate from three primary sources:
(1) faults in the MCP Software Development Kits (SDKs),
(2) faults in the underlying LLM models, and
(3) faults in the MCP server code itself.
This study focuses exclusively on faults originating from the MCP server code.

\section{Methodology}
\label{sec:method}
\begin{figure}
    \centering
    \includegraphics[width=\linewidth]{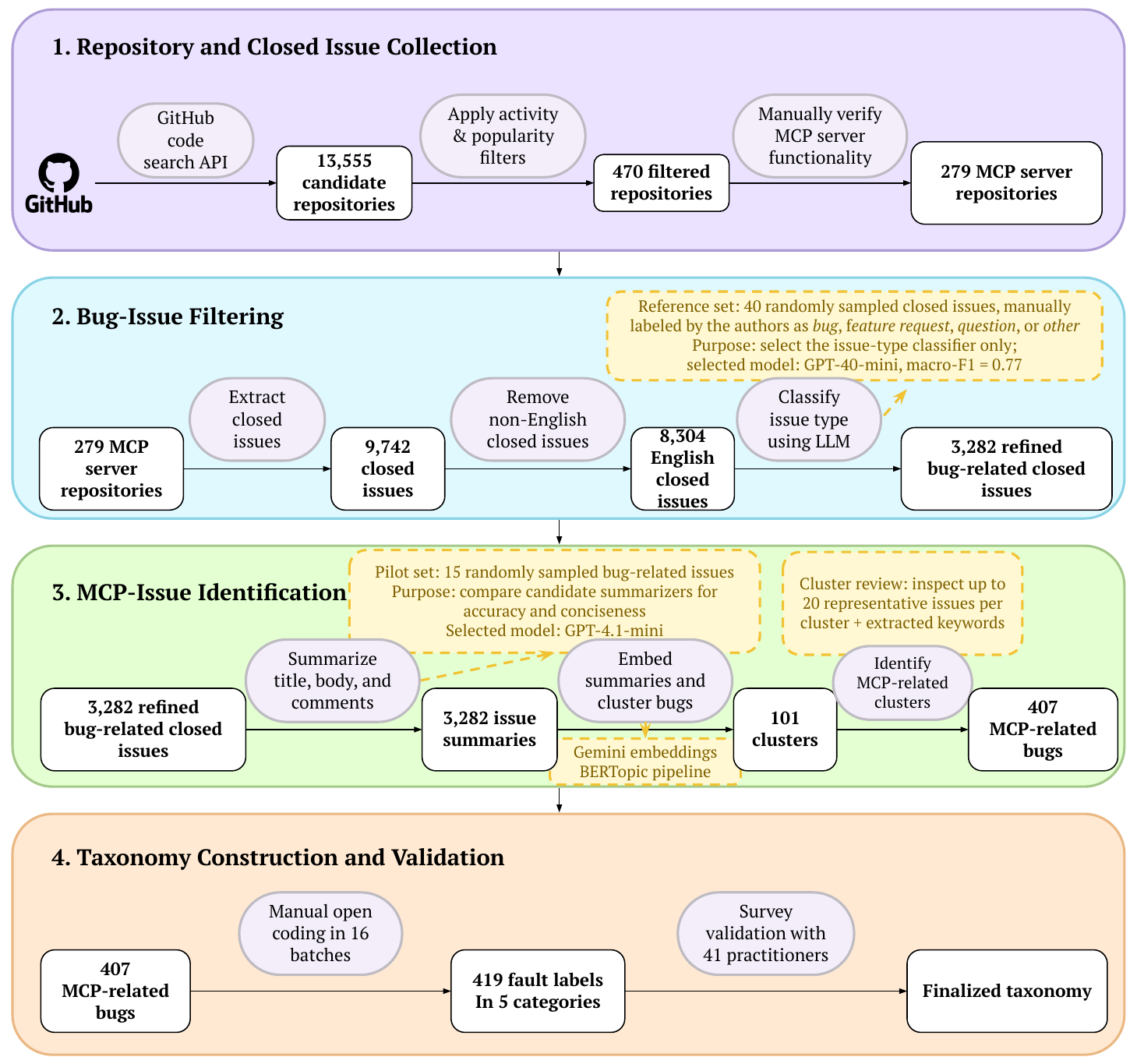}
    \caption{Overview of the used methodology}
    \label{fig:method}
\end{figure}

This section presents the methodology used to investigate bugs in MCP servers.
Fig.\ref{fig:method} presents a high-level view of the methodology followed in this study.
The primary source of data for our analysis is GitHub. With over 150 million registered users and more than 420 million repositories as of August 2025, GitHub is widely regarded as the most significant platform of open-source software systems in computer society~\cite{aghili2023studying}.

\subsection{Collecting Repositories}
\label{collect_repos}
Since standard Software Development Kits (SDKs) are available for multiple programming languages to implement MCP servers, this study focuses on repositories that rely on these SDKs. 
Given that Python is widely recognized as the most popular programming language for developing AI-based software systems~\cite{morovati2024bug,morovati2024fault}, we concentrate on repositories that employ the MCP Python SDK in their applications.
To collect the required data, we leverage the GitHub Search API~\cite{github_api_v3}, which allows us to extract detailed information about repositories. Specifically, we search for repositories that apply MCP in their provided software by identifying the use of the MCP library, including all possible import statements showing the usage of MCP (such as \texttt{‘import mcp’}, \texttt{‘from mcp import’}, \texttt{‘from mcp.server import’}, etc).
It is important to note that GitHub restricts access to only the first \num{1000} results returned by the search API. To overcome this limitation, we execute 
multiple search queries by constraining the size of searched files within defined ranges (using the parameter `\texttt{size:<min>..<max>}') to ensure that each query returned fewer than \num{1000} results. Since the search API do not return any results for file sizes greater than 350 KB in our case, we divide the entire range from 1 B to 350 KB into intervals of 50 B. This process results in a total of approximately 7,000 API calls.
We then extract the unique repositories containing code related to MCP (achieved from previous step), identifying \num{13555} repositories implementing an MCP server.

In the next step, we apply a set of exclusion criteria to examine the metadata of the extracted repositories and filter out unpopular ones. The exclusion criteria adopted in this study are consistent with those successfully employed in prior work~\cite{humbatova2020taxonomy,morovati2023bugs}.

\begin{itemize}
    \item Repositories with fewer than 10 stars and 10 forks 
    \item Repositories without any closed issues after November 1, 2024, since MCP has been officially introduced in November 2024 by Anthropic~\cite{mcp_introduction}. 
\end{itemize}

After applying the initial exclusion criteria, we identified 470 repositories implementing MCP server systems. We then manually reviewed all repositories to confirm the presence of MCP-specific functionality. For this verification, we applied an additional set of exclusion criteria, following the approach of prior studies~\cite{morovati2024bug}, to perform a final round of filtering.

\begin{itemize}
    \item Repositories with the description's language other than English (e.g. \cite{yzfly_douyin-mcp-server_2026, kotaro-kinoshita_yomitoku_2026})
    \item Repositories only containing tutorials and training materials, which can be a set of sample codes (e.g. \cite{nirdiamant_genai_agents_2026, azure-samples_semantic-kernel-advanced-usage_2025}).
    \item Repositories that use MCP solely for testing implemented functionalities, rather than for developing software features leveraging MCP capabilities (e.g. \cite{hazyresearch_minions_2026,llamastack_llama-stack_2026}).
\end{itemize}

At this stage, we obtain 279 repositories implementing MCP servers. We then extracted all closed issues from these repositories, resulting in 9,742 issues.
Since some issues have been reported in languages other than English, even within repositories that provide complete English documentation (such as \cite{noEnglish_ex2,noEnglish_ex3}),
we verified the language used in issue descriptions by applying Python Regular Expressions (regex)~\cite{python_regex}. Issues written in non-English languages were removed. After this filtering, we obtained 8,304 closed issues associated with MCP server repositories.

\subsection{Labeling Issues}

This section outlines the methodology used to label the extracted closed issues from MCP-based repositories in order to construct the taxonomy. First, we classify issues to distinguish bugs from other types of reported issues. Next, we apply clustering techniques to group conceptually related closed issues
Finally, we focus on clusters with a direct relationship to MCP functionality and manually inspect all issues.

\subsubsection{Classifying Closed Issues}
\label{subsubsec:classification}

Our classification process first distinguishes bug reports from all other forms of closed issues. We perform this classification step beyond relying on GitHub labels since GitHub labels are project-level metadata rather than a standardized classification scheme that is consistently applied across repositories. Assigning GitHub labels can be optional and issues may be associated with multiple labels simultaneously~\cite{nadeem2021automatic}, while individual projects frequently define custom labels to encode different aspects of development and issue-management information~\cite{kim2021empirical}. Recent work on automatic issue classification further identifies label correctness and data quality as construct-validity concerns~\cite{colavito2024impact}. Consequently, directly using GitHub labels may lead to inconsistencies and reduce the comparability of issue categories across projects. We therefore apply a standardized issue-type classification scheme to the collected closed issues. Overall, GitHub issues are typically categorized into three primary types: \textit{bug reports}, \textit{feature requests}, and \textit{questions}~\cite{kallis2021predicting,siddiq2022bert}. Beyond these established categories, we introduce an additional class, denoted as \textit{Other}, to capture closed issues that do not clearly fall into any of the three conventional types. This additional category avoids forcing such issues into an unsuitable category.

To automate this classification across the entire closed-issue corpus, we consider several widely used open-source and proprietary LLMs as candidate classifiers. To evaluate these candidates, we construct an author-annotated reference set by randomly sampling 40 issues from the collected closed-issue corpus. The first two authors independently assign each reference issue one of the four issue-type labels. Their initial annotations achieve a Cohen’s kappa of 0.75, and they resolve disagreements through discussion to determine the final labels. The resulting author-assigned labels, rather than the existing GitHub labels, serve as the ground truth for evaluating the candidate models. The reference set is used only to select the issue-type classifier and is not used to construct or validate the MCP fault taxonomy. This use is consistent with empirical software-engineering studies that employ pilot subsets to calibrate data-labeling or extraction procedures before the main analysis~\cite{zhou2025exploring}.

To evaluate the proprietary OpenAI models on the author-annotated reference set, we use the OpenAI API Python library \cite{openai_python_2025} and the Chat Completions API. Table~\ref{tbl:LLMs} reports the exact OpenAI model snapshots evaluated. For the open-source candidates, including Llama, Mistral, Gemma, Qwen, and Zephyr, we rely on their corresponding Ollama implementations~\cite{ollama:ollama2024}. We execute these models using the quantized formats provided by Ollama. Llama~\cite{ollama:ollama}, Qwen~\cite{ollama_qwen}, and Gemma~\cite{ollama_gemma} run in the \texttt{Q4\_K\_M} format, while Mistral~\cite{ollama_mistral} and Zephyr~\cite{ollama_zephyr} run in the \texttt{Q4\_0} format. We report the quantization settings to ensure reproducibility of our evaluation setup.

All candidate models receive the same classification prompt. The prompt consists of a fixed system instruction defining the issue-type categories and a user message containing the issue title, description, and comments. For each reference issue, every candidate model is prompted to assign exactly one of the four issue-type labels (Appendix \ref{app:prompt}, Listing ~\ref{lst:prompt-classification1}). For the GPT-4.1, GPT-4.1-mini, and GPT-4o-mini candidates, we set the temperature to 0.2 to reduce stochastic variation and encourage stable outputs. We use the same temperature for the Ollama models. Because o4-mini does not support changes to temperature, we evaluate it using its default decoding configuration. This use of a low temperature is consistent with recent LLM-based software-engineering classification studies~\cite{chaudhary_exploring_2025, oztas_towards_2025}. For the OpenAI models, we do not explicitly set top-$p$, maximum output tokens, random seed, frequency or presence penalties, or stop sequences. Supported parameters therefore retain their API defaults.

Because the distribution of issue types in the reference set is imbalanced, we report macro-averaged precision, recall, and F1-score, computed using the \textit{scikit-learn} library~\cite{pedregosa2011scikit}. Macro-averaged metrics weight all classes equally and provide a more appropriate comparison when some categories are underrepresented~\cite{hinojosa2024performance}. Table~\ref{tbl:LLMs} presents the performance of the candidate models.

\begin{table}[]
\small
\caption{\centering Performance of the candidate LLMs on the author-annotated reference set of 40 closed issues (AC = \textit{Accuracy}, PR = macro-averaged \textit{Precision}, RC = macro-averaged \textit{Recall}, and F1 = macro-averaged \textit{F1-score}).}
    \centering
    \begin{tabular}{p{6.5cm} r r r r}
        \hline
        \multirow{2}{6em}{Models}
        & \multicolumn{4}{c}{Issue Classification Results} \\
        \cline{2-5}
        & \multicolumn{1}{c}{\textbf{\textit{AC}}}
        & \multicolumn{1}{c}{\textbf{\textit{PR}}}
        & \multicolumn{1}{c}{\textbf{\textit{RC}}}
        & \multicolumn{1}{c}{\textbf{\textit{F1}}} \\
        \hline
        Llama3.1:8b \cite{Llama-31-8B} & 0.77 & 0.81 & 0.82 & 0.76 \\
        Gemma3:4b \cite{gamma-7b} & 0.69 & 0.58 & 0.58 & 0.50 \\
        zephyr:7b \cite{zephyr-7b} & 0.67 & 0.62 & 0.51 & 0.53 \\
        Mistral:7b \cite{mistral-7B} & 0.67 & 0.70 & 0.58 & 0.60 \\
        Qwen3:8b \cite{qwen3-8B} & 0.69 & 0.61 & 0.59 & 0.59 \\
        GPT-4.1 (\texttt{gpt-4.1-2025-04-14}\cite{openai_gpt41}) & 0.74 & 0.76 & 0.81 & 0.75 \\
        GPT-4o-mini (\texttt{gpt-4o-mini-2024-07-18}\cite{gpt-4o-mini}) & 0.77 & 0.78 & 0.82 & \textbf{0.77} \\
        o4-mini (\texttt{o4-mini-2025-04-16}\cite{openai_o4mini}) & 0.64 & 0.45 & 0.55 & 0.44 \\
        GPT-4.1-mini (\texttt{gpt-4.1-mini-2025-04-14}\cite{openai_gpt41mini}) & 0.74 & 0.79 & 0.81 & 0.74 \\
        \hline
    \end{tabular}
    \label{tbl:LLMs}
    \vspace{-1em}
\end{table}
GPT-4o-mini achieves the highest macro-averaged F1-score of 0.77 and is therefore selected as the issue-type classifier. The selected classifier's macro-F1 is also consistent with prior cross-project issue-classification studies, which report F1 or macro-F1 values in the high-0.7 to low-0.8 range under realistic imbalance, multi-label, or label-quality constraints ~\cite{kallis2021predicting,nadeem2021automatic,aracena2024applying,colavito2024impact,colavito2024leveraging}. We apply the selected model to the complete closed-issue corpus. Issues classified as bug reports proceed to the filtering step described below.

\begin{table}
\small
\caption{\centering Detailed information about the number of remaining repositories/closed issues 
after each filtering step.
}
    \centering
    \begin{tabular}{p{6cm} r }
        \hline
            \textbf{Filtering Step} &  \multicolumn{1}{c}{\textbf{\textit{\# items}}} \\ 
            \hline
            All extracted repositories & \num{13555}   \\
            Repositories after removing unpopular & \num{470} \\
            Repositories after manual checking &  \num{279}\\
            \hline
            All extracted closed issues &  9,742 \\
            Closed issues after removing non-english & 8,304 \\
            Bug-related closed issues & 3,591 \\
            Bug-related closed issues after refinement & 3,282 \\
            \hline
    \end{tabular}
        \label{tbl:repo_filter}
        \vspace{-1em}
\end{table}

Bug-related issues that are closed due to being stale, duplicates, or not reproducible are generally unsuitable for empirical analysis, as they do not reflect actionable or validated defects~\cite{li2022follow}. Accordingly, such issues must be excluded from our dataset to ensure the validity of subsequent analyses.
To perform this filtering, we use GPT-4o-mini (selected in the previous classification step) and prompt it to identify bug-related issues closed as stale, duplicate, or non-reproducible (Appendix~\ref{app:prompt}, Listing~\ref{lst:prompt-classification2}). 
Among the 3,591 bug-related issues, 309 issues meeting at least one of these criteria are excluded from the dataset, resulting in a refined set of 3,282 bug-related issues. 
For instance, issue~\cite{issue_stale} is removed during this phase because it was closed as stale. This set of 3,282 bug-related issues is retained for the subsequent clustering and manual taxonomy-construction stages.
Table~\ref{tbl:repo_filter} provides detailed information on the remaining repositories and closed issues at each stage of the filtering process.

\subsubsection{Clustering Bugs}\label{subsubsec:clustering}

It is important to note that the 3,282 bug-related closed issues
span a broad range of software components, including user interfaces, web interfaces, and database systems. However, our objective is to focus on those issues specifically related to MCP components. To achieve this, we cluster the bug-related issues to identify topic-based groups and extract the subset corresponding to MCP servers. For clustering, we employ the BERTopic framework~\cite{grootendorst_maartengrbertopic_2025}, a modular topic modeling approach that integrates transformer-based embeddings, dimensionality reduction, and density-based clustering to produce semantically coherent document groups~\cite{grootendorst_bertopic_2022}. 

Given that issue comments provide complementary insights into the nature of reported issues~\cite{ramirez2020descriptions}, the true characteristics of a closed issue often become clearer when the issue title and description are analyzed alongside the subsequent discussion, including the comment that ultimately leads to its closure. Accordingly, consistent with prior work~\cite{choetkiertikul2024sprint2vec,alam2026analyzing}, we construct the textual representation of each issue by aggregating its title, body, and full comment thread.
However, issue comments vary considerably in length, may contain extended discussions among contributors, and can occasionally diverge from the central topic. These factors may negatively affect clustering quality. To mitigate this issue, we first generate a concise summary of each closed issue (including title, body, and comments) using LLMs, and subsequently perform topic modeling on the generated summaries.

We evaluate five candidate LLMs for the summarization task, including four GPT variants (Table~\ref{tbl:LLMs}) and \texttt{Llama3.1:8b}. The inclusion of \texttt{Llama3.1:8b} is motivated by its competitive performance relative to leading GPT models in the preceding classification task (Table~\ref{tbl:LLMs}).
To compare summarization quality, we randomly select a sample of 15 bug-related closed issues. 
It is important to note that similar to the 40-issue reference set used for classifier selection, this pilot sample is used exclusively to choose the summarization model for preprocessing of our dataset, not to derive taxonomy categories, validate the final taxonomy, or estimate the prevalence of MCP faults. This design is consistent with LLM summarization evaluation work in which a small set of documents is manually inspected against the source text~\cite{dehkordi2025improving}, and with empirical software-engineering studies that use pilot subsets to calibrate analysis procedures before the main analysis~\cite{zhou2025exploring}.
Each model generates summaries using the same prompt (Appendix~\ref{app:prompt}, Listing~\ref{lst:prompt-summarization}). The first two authors independently evaluate the summaries with respect to accuracy and conciseness. Accuracy refers to whether the summary faithfully preserves the issue's main problem, context, and resolution evidence from the title, body, and comments. Conciseness also refers to whether the summary captures that information in the requested 1--2 sentence format without adding irrelevant details. Disagreements are subsequently resolved through discussion.
The evaluation results indicate that \texttt{GPT-4.1-mini} produces the best summaries in 9 out of 15 cases, while \texttt{GPT-4.1} performs best in 4 cases. For two instances where the raters initially disagreed (between \texttt{GPT-4.1-mini} and \texttt{GPT-4.1}), consensus was reached in favor of \texttt{GPT-4.1-mini}. Overall, \texttt{GPT-4.1-mini} demonstrates superior performance in 11 out of 15 samples, compared to 4 out of 15 for \texttt{GPT-4.1}. Considering both performance and cost efficiency, we select \texttt{GPT-4.1-mini} as the summarization model for the full dataset.
For the full-dataset summarization step, we use the fixed model snapshot \texttt{gpt-4.1-mini-2025-04-14} through the OpenAI Chat Completions API. The prompt uses a fixed format that asks the model to produce a concise \texttt{Subject} and a one- to two-sentence \texttt{Summary} from the issue title, body, and comments. As in the classification step, we set the temperature value of 0.2, to achieve stable and reproducible outputs for preprocessing, and we do not explicitly set top-$p$, maximum output tokens, random seed, frequency or presence penalties, or stop sequences; these parameters therefore follow the OpenAI API defaults.

Next, we use precomputed semantic embeddings of the closed-issue summaries as input to the BERTopic pipeline. To obtain these embeddings, we select the \texttt{gemini-embedding-001} model~\cite{lee_gemini_2025, gemini_embeddings_nodate}, which ranks among the top-performing models on the \textit{Massive Text Embedding Benchmark (MTEB)} leaderboard~\cite{enevoldsen_mmteb_2024, hf_mteb_nodate}. The model generates 3,072-dimensional dense vector representations that capture contextual and semantic relationships between issue summaries.
Direct clustering in such high-dimensional spaces is both computationally expensive and statistically challenging due to the well-known curse of dimensionality, where distance metrics become less discriminative and data points tend to appear equidistant~\cite{aggarwal2015data}. To mitigate these effects and improve cluster separability, we apply Uniform Manifold Approximation and Projection (UMAP)~\cite{mcinnes_umap_2020} to reduce the embedding dimensionality to five. UMAP is a non-linear manifold learning technique that preserves local neighborhood structure while maintaining meaningful global relationships, making it particularly suitable for downstream clustering tasks on semantic embeddings.
Following dimensionality reduction, we apply \textit{KMeans} clustering~\cite{sinaga2020unsupervised} to partition the summaries into 101 clusters. KMeans is chosen for its computational efficiency, stability, and suitability for partitioning compact, low-dimensional representations~\cite{2012:DataMining}. To determine an appropriate number of clusters ($k$), we first apply the density-based \textit{HDBSCAN} algorithm~\cite{mcinnes_hdbscan_2017}, which identifies the intrinsic grouping structure of the data without requiring a predefined cluster count. The resulting density-based structure informs our selection of $k$, enabling a balance between cluster granularity and interpretability.

To interpret the resulting clusters, BERTopic applies a CountVectorizer followed by the class-based Term Frequency-Inverse Document Frequency (c-TF–IDF) weighting scheme to extract the most representative terms per cluster. We also compute a two-dimensional UMAP projection of the embeddings to visualize the discovered clusters and later construct a hierarchical topic tree following BERTopic’s hierarchical topic modeling procedure. 

In this stage, we analyze the bug-related closed issues within each cluster to identify those directly related to MCP components. A cluster is considered MCP-related if its issues affect the functionality of core MCP components (e.g., MCP server, MCP tools, or MCP host).
Consistent with established topic-modeling practice, where topics are interpreted through representative terms and human inspection~\cite{costasilva2021topic,weston2023selecting}, we use two complementary keyword signals to support cluster interpretation: BERTopic's top c-TF--IDF terms and an auxiliary set of discriminative keywords extracted from issue summaries. The auxiliary extraction normalizes summaries, removes English and domain-generic stopwords, considers unigram and bigram terms, and ranks terms for each cluster using TF--IDF with one-vs-rest chi-square scores.
To make the MCP-relevance decision reproducible, the first two authors (one Ph.D. researcher and one Ph.D. candidate with practical expertise in MCP) independently examine these keyword signals and manually review up to 20 randomly selected issues per cluster, inspecting all issues for clusters with fewer than 20 issues. Given that the median cluster size is 28, this review covers a substantial portion of a typical cluster. The sampled issues are used to determine whether the cluster's dominant theme concerns MCP protocol, server, tool, or host behavior. We label a cluster as MCP-related when both raters agree that at least 60\% of the inspected issues in the cluster concern MCP behavior and that the keyword signals are consistent with this interpretation.
Following prior methodological practices~\cite{morovati2024bug}, the raters begin by labeling the first 10 clusters (10\% of all clusters) to calibrate their understanding of the labeling criteria. This initial phase results in a Fleiss’ kappa agreement of 1.00~\cite{falotico2015fleiss}, indicating almost perfect agreement~\cite{instruments2012validity}.
They subsequently label the remaining clusters independently. After annotating every 30 clusters, they meet to review and resolve any disagreements. If consensus cannot be reached, a third rater adjudicates and makes the final determination regarding whether the cluster is related to MCP.
The overall inter-rater agreement is 0.94 using Fleiss’ kappa, indicating a high level of inter-rater reliability.
As a result, 9 out of 101 clusters are identified as MCP-related, encompassing 383 closed issues. 
Additionally, from the remaining clusters, we include any issue whose generated summary contains the keyword \texttt{`mcp'}, using case-insensitive matching, yielding a final dataset of 407 MCP-related closed issues. This keyword matching step is used only as a post-processing step to capture MCP-related issues that were not included through cluster-level selection.
These issues are then aggregated and reserved for detailed manual labeling in subsequent analysis.

For the remaining 92 non-MCP clusters, we assign descriptive labels summarizing their dominant themes. Consistent with established practices in topic modeling studies within software engineering research~\cite{li2021understanding,alam2025empirical}, we interpret each cluster by analyzing its top-ranked keywords and manually reviewing 20 randomly selected representative issues. Based on this analysis, we assign concise and meaningful labels that capture the primary focus of each cluster.
Subsequently, we merge clusters with overlapping themes and reorganize them into a cleaner set of categories to improve interpretability. A high-level overview of the identified categories and their respective sizes is presented in Table~\ref{tbl:clustering}.

\begin{table}[]
\footnotesize 
\caption{\centering Overview of high-level bug topics derived from clustering}
\label{tbl:clustering}
\begin{tabular}{lr}
\hline
\textbf{Category}                       & \multicolumn{1}{l}{\textbf{Size}} \\ \hline
\rowcolor[HTML]{EFEFEF} 
UI and UX                               & 457                               \\
Repo-specific integration/configuration & 435                               \\
\rowcolor[HTML]{EFEFEF} 
\textbf{MCP-specific Issues}            & \textbf{383}                      \\
Database and Data Handling              & 310                               \\
\rowcolor[HTML]{EFEFEF} 
LLM-Specific                            & 300                               \\
Environment and Dependency              & 224                               \\
\rowcolor[HTML]{EFEFEF} 
Deployment                              & 177                               \\
File path and URL handling              & 142                               \\
\rowcolor[HTML]{EFEFEF} 
Web (HTML/CSS)                          & 141                               \\
API configuration                       & 110                               \\
\rowcolor[HTML]{EFEFEF} 
Agent-Specific                          & 77                                \\
Authentication and Authorization        & 70                                \\
\rowcolor[HTML]{EFEFEF} 
Runtime and Startup Failures            & 70                                \\
Data Type and Schema Handling           & 66                                \\
\rowcolor[HTML]{EFEFEF} 
Concurrency and Synchronization         & 64                                \\
File Format and Encoding                & 60                                \\
\rowcolor[HTML]{EFEFEF} 
Documentation                           & 53                                \\
Testing and Continuous Integration (CI)                             & 48                                \\
\rowcolor[HTML]{EFEFEF} 
Workflow and Orchestration              & 43                                \\
Build                                   & 25                                \\
\rowcolor[HTML]{EFEFEF} 
Other                                   & 27                                \\ \hline
\textbf{Total}                          & \multicolumn{1}{l}{\textbf{3,282}} \\ \hline
\end{tabular}
\end{table}

\subsubsection{Manual annotation of MCP-related Bugs}\label{subsubsec:manual_labeling}

To construct a taxonomy of faults specific to MCP-based applications, we manually inspect 407 bug-related closed issues belonging to MCP-related clusters. The first two authors 
independently label the issues using an open coding procedure~\cite{seaman1999qualitative}. 
Since no predefined labels are available, we cannot calculate inter-rater agreement in this step~\cite{Saldana2021}. 
We divide the dataset into 16 parts (15 batches with 25 issues and 1 batch with 32 issues),
allowing the raters to meet after labeling each batch to discuss their coding decisions, review emerging categories, and refine or merge them, following the methodology of prior studies~\cite{humbatova2020taxonomy,morovati2024bug}.
Once all MCP-related issues are coded, the raters meet again to consolidate the final set of labels, resolve disagreements, and produce a codebook. 
In cases where the two primary raters are unable to resolve a disagreement, a third author is consulted to adjudicate and make the final decision regarding the appropriate label for the issue.
Finally, the raters re-examined all closed issues to ensure that the assigned labels were consistent with the finalized codebook. We considered saturation to be reached at batch 10, as no new categories emerged in two consecutive batches.
It is also worth noting that some issue reports describe two or more MCP-related faults, and were therefore assigned multiple labels accordingly. 
The final MCP-related fault categories, along with their corresponding definitions, are presented in Section~\ref{subsec:taxonomy_results}. The complete codebook is available in our replication package~\cite{replication_package}.

\subsection{Taxonomy Creation and Validation}
Following prior studies~\cite{vijayaraghavan2003bug,morovati2024common}, we apply a bottom-up methodology to construct the taxonomy of faults. After labeling all MCP-related closed issues and developing the final codebook, we group labels with similar themes into categories. Then, we organize these categories and their subcategories into a hierarchical model, ensuring that the relationships followed an `is-a' structure. Finally, all authors carefully review the complete taxonomy to validate its accuracy and finalize it.

To evaluate the comprehensiveness and representativeness of the proposed taxonomy, we conducted a survey with researchers and practitioners. Surveys enable the systematic elicitation of expert judgments at scale~\cite{shackleton2021interviews,aldhaen2020interview}, and are well suited for assessing the completeness and relevance of conceptual frameworks. The results provide empirical evidence supporting the taxonomy’s external validity and practical applicability~\cite{nekkanti2016surveys}.

The survey targets MCP practitioners who were not involved in the taxonomy construction process, thereby reducing confirmation bias. To identify potential participants, we extracted contributors from the MCP server repositories identified in the previous phase of our study. Following established approaches in mining software repositories~\cite{morovati2024common}, we used the GitHub REST API v3 to collect contributor information. Because the survey was primarily distributed via email~\cite{ghazi2018survey}, we retained only practitioners who publicly shared their email addresses in their profiles. This procedure resulted in 10,756 unique developer email addresses.

We designed and administered the survey using Qualtrics~\cite{survey_ualtrics}, a widely adopted platform for academic research~\cite{velykoivanenko2024designing,humbatova2020taxonomy}. In addition to direct email invitations, we disseminated the survey through online forums (e.g., Discord communities) dedicated to MCP development to broaden participation.
 
The survey begins with general questions about participants’ current roles and the programming languages they use for developing MCP servers. Since MCP had been publicly available for less than a year when we administered the survey, we did not include questions about prior MCP server development experience.
The next section focuses on the proposed taxonomy. To avoid overwhelming participants with a single, complex figure, we organized the questions into four groups corresponding to the high-level MCP-specific fault categories in the taxonomy. We exclude \textit{General Programming} from the survey, because that category captures conventional implementation faults that are not specific to MCP software. Each fault type is accompanied by a brief description to ensure a clear understanding of the concept.
For each fault type, participants are first asked whether they have encountered the fault (\textit{yes/no}). Two follow-up questions assess the fault’s severity (\textit{minor, major, critical}) and the effort required to resolve it (\textit{low, medium, high}). Finally, participants are invited to report any MCP-related faults not captured by the taxonomy, helping to identify potential omissions. The full survey instrument is available in the study’s replication package~\cite{replication_package}.

\section{Results}
\label{sec:results}
This section presents the results of our study. We first summarize the main findings across the three RQs and then provide the detailed taxonomy, survey validation, and fault-characteristic analyses. All data used in the analysis are publicly available at~\cite{replication_package}.

\begin{mainfindingsbox}{Main Findings Across RQs}
\begin{itemize}
    \item \textbf{F1. Boundary concentration:} MCP faults are concentrated at component boundaries where servers interact with hosts, server-defined tools, invoked resources, runtime environments, and protocol-level configuration.
    \item \textbf{F2. Response-boundary failures:} Practitioners report \textit{tool response handling} as the most frequently encountered fault, showing that failures can occur even after tool discovery and invocation succeed.
    \item \textbf{F3. Divergent prioritization signals:} Fault prevalence, severity index, critical-response share, and effort index point to different fault types, indicating that prioritization should consider multiple evidence signals rather than issue frequency alone.
    \item \textbf{F4. Host-side diagnostic friction:} Host-related faults often require developers to reason about user-specific launch commands, configuration files, transport choices, logs, and execution environments.
    \item \textbf{F5. Discussion depth:} MCP-related faults generate more intensive diagnostic discussion than non-MCP faults, reflecting the need to reason across protocol interactions, returned tool payloads, host settings, and user environments.
\end{itemize}
\end{mainfindingsbox}

To make the connection between the main findings, their supporting evidence, and the actionable insights explicit, Table~\ref{tbl:findings_implications} summarizes the traceability across these elements.

\begin{table}[htbp]
\small
\caption{Summary linking major findings, supporting evidence, and actionable implications. IDs F1--F5 refer to the main findings summarized above, and IDs A1--A5 refer to the actionable insights discussed in Section~\ref{subsec:implications}.}
\label{tbl:findings_implications}
\centering
\resizebox{\textwidth}{!}{
\begin{tabular}{L{3.1cm} L{5.3cm} L{6.2cm}}
\toprule
\textbf{Finding} & \textbf{Supporting Evidence} & \textbf{Developer Action} \\
\midrule
\textbf{(F1)} Boundary concentration &
\textit{Server/Tool Configuration} (31.74\%), \textit{Server/Host Configuration} (28.64\%), \textit{Server Setting} (27.45\%) &
\textbf{(A1)} Test host--server--tool launch paths, transport modes, configuration templates, working directories, and environment assumptions \\
\midrule
\textbf{(F2)} Response-boundary faults &
\textit{Tool response handling} has the highest survey prevalence (66.67\%) &
\textbf{(A2)} Validate response contracts, size limits, serialization, structured errors, and host-side parsing after tool invocation \\
\midrule
\textbf{(F3)} Divergent signals &
\textit{Host connection mismatch} has the highest severity index; \textit{tool discovery \& registration} has the highest critical-response share; \textit{tool authentication} has the highest effort index &
\textbf{(A5)} Prioritize release-gating and design-review effort using prevalence, severity, and repair effort together \\
\midrule
\textbf{(F4)} Host-side diagnosis &
\textit{Server/host configuration} has more comments and comments per collaborator than \textit{Documentation} &
\textbf{(A3)} Capture host version, transport mode, launch command, redacted config, logs, and minimal failing responses \\
\midrule
\textbf{(F5)} Discussion depth &
MCP faults have more comments and comments per collaborator than non-MCP faults, but not significantly more collaborators &
\textbf{(A4)} Reduce diagnostic iteration by making parsing, logging, transport, and protocol-channel failures explicit and reproducible \\
\bottomrule
\end{tabular}
}
\end{table}

\subsection{The Final Taxonomy (RQ1)}\label{subsec:taxonomy_results}
Figure~\ref{fig:mcp-taxonomy} provides an overview of the final taxonomy, derived through our manual labeling process. Each issue was examined and labeled, with some issues receiving multiple labels, resulting in a total of 419 labels. Twenty-four issues were excluded because their authenticity or root cause could not be determined.
\definecolor{cRoot}{HTML}{111111}
\definecolor{cServer}{HTML}{3F4A1E}
\definecolor{cGeneral}{HTML}{C00000}
\definecolor{cTool}{HTML}{1F3A6D}
\definecolor{cDoc}{HTML}{8B4A12}
\definecolor{cHost}{HTML}{8A6A00}

\definecolor{cLightServer}{HTML}{bfe89b}
\definecolor{cLightTool}{HTML}{b2daf7}
\definecolor{cLightHost}{HTML}{f7f399}

\definecolor{cMidServer}{HTML}{4f9c4e}
\definecolor{cMidTool}{HTML}{1886d6}
\definecolor{cMidHost}{HTML}{b8b104}

\forestset{
  paperbox/.style={
    draw,
    rounded corners=2pt,
    grow'=east, 
    align=center,
    font=\sffamily\small,
    inner sep=2pt,
    minimum height=5mm,
    text width=48mm,
    text=white,
    drop shadow,
    child anchor=west,
    parent anchor=east,
    edge path={
      \noexpand\path [draw, \forestoption{edge}] (!u.parent anchor) -- +(5pt,0) |- (.child anchor)\forestoption{edge label};
    },
  },
  rootbox/.style={
    paperbox,
    font=\sffamily\small,
    text width=17mm,
  },
  catbox/.style={
    paperbox,
    text width=26mm,
    minimum height=10mm,
  },
  leafbox/.style={
    paperbox,
    text width=40mm,
    font=\sffamily\scriptsize,
    text=black,
  },
  edge/.style={-latex, line width=0.45pt},
  mcpdiagram/.style={
    for tree={
      edge,
      grow'=east,
      s sep=3mm,
      l sep=5mm,
      drop shadow,
      child anchor=west,
      parent anchor=east,
      edge path={
        \noexpand\path [draw, \forestoption{edge}] (!u.parent anchor) -- +(5pt,0) |- (.child anchor)\forestoption{edge label};
      },
    },
  },
}

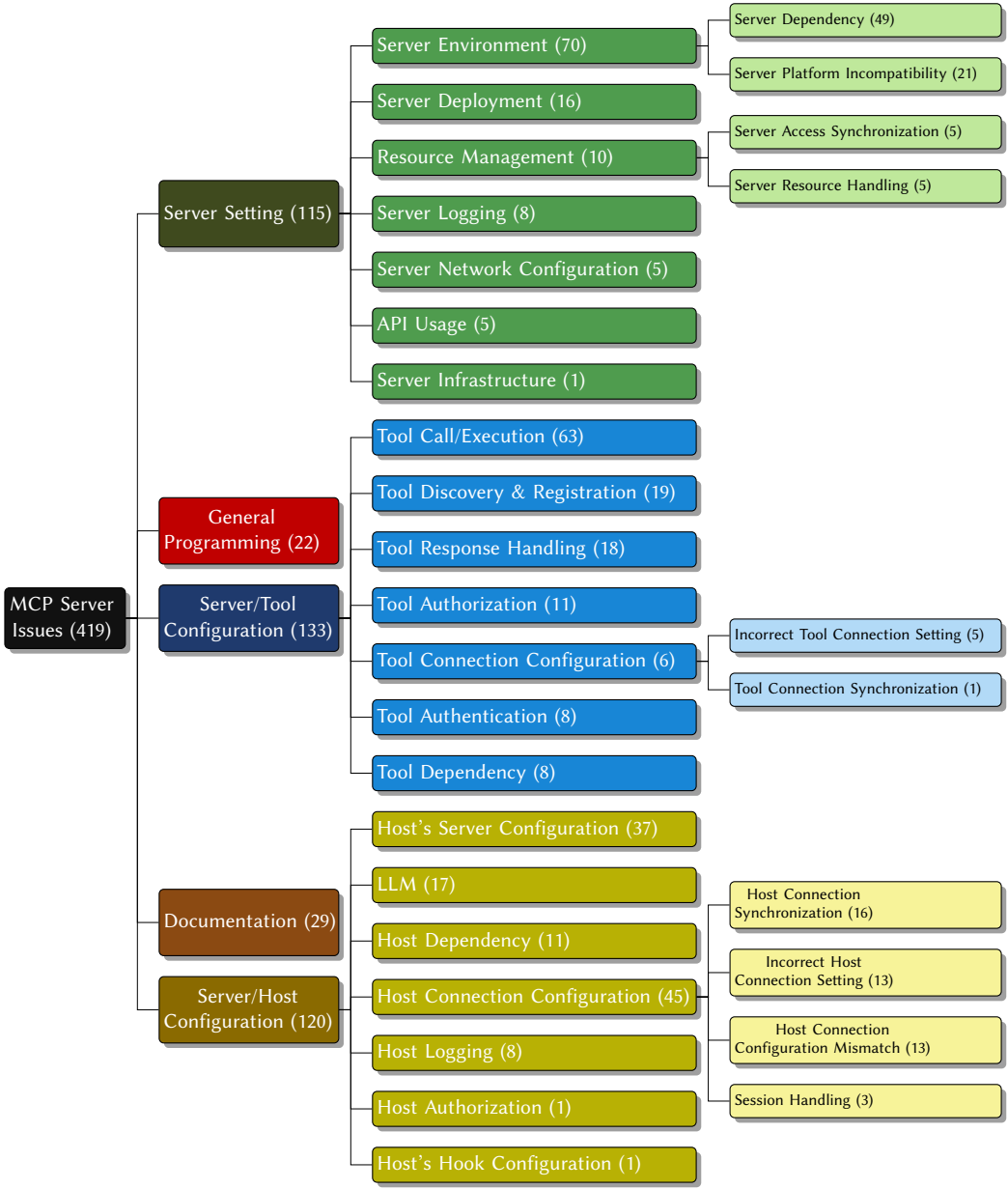
\begin{figure}[t]
  \centering
  \resizebox{\columnwidth}{!}{
    \begin{forest} mcpdiagram
      [
        {MCP Server\\Issues (419)},
        rootbox, fill=cRoot
        [
          {Server Setting (115)}, catbox, fill=cServer
          [{Server Environment (70)}, paperbox, fill=cMidServer
            [{Server Dependency (49)}, leafbox, fill=cLightServer]
            [{Server Platform Incompatibility (21)}, leafbox, fill=cLightServer]
          ]
          [{Server Deployment (16)}, paperbox, fill=cMidServer]
          [{Resource Management (10)}, paperbox, fill=cMidServer
            [{Server Access Synchronization (5)}, leafbox, fill=cLightServer]
            [{Server Resource Handling (5)}, leafbox, fill=cLightServer]
          ]
          [{Server Logging (8)}, paperbox, fill=cMidServer]
          [{Server Network Configuration (5)}, paperbox, fill=cMidServer]
          [{API Usage (5)}, paperbox, fill=cMidServer]
          [{Server Infrastructure (1)}, paperbox, fill=cMidServer]
        ]
        [
          {General\\Programming (22)}, catbox, fill=cGeneral
        ]
        [
          {Server/Tool\\Configuration (133)}, catbox, fill=cTool
            [{Tool Call/Execution (63)}, paperbox, fill=cMidTool]  
            [{Tool Discovery \& Registration (19)}, paperbox, fill=cMidTool]
            [{Tool Response Handling (18)}, paperbox, fill=cMidTool]
            [{Tool Authorization (11)}, paperbox, fill=cMidTool]
            [{Tool Connection Configuration (6)}, paperbox, fill=cMidTool
                [{Incorrect Tool Connection Setting (5)}, leafbox, fill=cLightTool]
                [{Tool Connection Synchronization (1)}, leafbox, fill=cLightTool]
            ]
            [{Tool Authentication (8)}, paperbox, fill=cMidTool]
            [{Tool Dependency (8)}, paperbox, fill=cMidTool]
        ]
        [
          {Documentation (29)}, catbox, fill=cDoc
        ]
        [
          {Server/Host\\Configuration (120)}, catbox, fill=cHost
            [{Host's Server Configuration (37)}, paperbox, fill=cMidHost]
            [{LLM (17)}, paperbox, fill=cMidHost]
            [{Host Dependency (11)}, paperbox, fill=cMidHost]
            [{Host Connection Configuration (45)}, paperbox, fill=cMidHost
                [{Host Connection \\Synchronization (16)}, leafbox, fill=cLightHost]
                [{Incorrect Host \\Connection Setting (13)}, leafbox, fill=cLightHost]
                [{Host Connection \\Configuration Mismatch (13)}, leafbox, fill=cLightHost]
                [{Session Handling (3)}, leafbox, fill=cLightHost]
            ]
            [{Host Logging (8)}, paperbox, fill=cMidHost]
            [{Host Authorization (1)}, paperbox, fill=cMidHost]
            [{Host's Hook Configuration (1)}, paperbox, fill=cMidHost]
        ]
      ]
    \end{forest}
    }
  \caption{Taxonomy of MCP server issues (counts in parentheses).}
  \label{fig:mcp-taxonomy}
\end{figure}

The labels are organized into five high-level categories: \textit{Server Setting} (27.45\%), \textit{Server/Tool Configuration} (31.74\%), \textit{Server/Host Configuration} (28.64\%), \textit{Documentation} (6.92\%), and \textit{General Programming} (5.25\%). Each category is further divided into a hierarchy of subcategories, capturing the diversity and specificity of MCP-related faults. We organize the taxonomy according to the architectural locus at which the fault is introduced or primarily resolved, rather than by recurring surface mechanisms alone. This choice reflects the layered structure of MCP systems: an MCP server must first be installed and executed in a suitable runtime environment, then expose and execute tools correctly, and finally interoperate with a host that launches, configures, and communicates with the server. Consequently, similar-looking mechanisms may appear in different high-level categories when their root cause and resolution belong to different MCP boundaries. For example, a dependency fault is classified under \textit{Server Setting} when the server itself cannot be installed or started because of missing or incompatible runtime packages, under \textit{Server/Tool Configuration} when a tool-specific dependency prevents a registered tool from loading or executing, and under \textit{Server/Host Configuration} when the host-side launch environment or host update causes dependency-related failures. This organization preserves the practical distinction between who can diagnose and fix the fault and where the corresponding mitigation should be applied.

To keep the Results section focused on the taxonomy structure and its interpretation, the following subsections provide concise category descriptions; representative issue examples illustrating the categories are reported separately in Appendix~\ref{app:taxonomy_examples}, Table~\ref{tbl:taxonomy_examples}.
\subsubsection{\textbf{Server Setting (115 issues)}}
\label{sec:server_setting}
This category encompasses issues arising from the configuration, setup, and operating context of the MCP server, problems that prevent the MCP server from starting, functioning reliably, or interacting properly with its execution environment. We classify these issues into seven subcategories:
\begin{enumerate}
    \item \textbf{Server Environment (70 issues)}: 
    Issues stemming from the configuration and compatibility of the server’s underlying runtime environment. These faults arise when the server cannot operate correctly due to missing or mismanaged dependencies, incompatible library or tool versions, or differences in Operating Systems (OS), architectures, or underlying platform environments. This subcategory comprises two groups:
        \begin{enumerate}
            \item \textit{Server Dependency (49 issues)}: Issues caused by missing, outdated, improperly installed, or incompatible dependencies required by the MCP server. These faults include absent Python modules, executables not available on the system \texttt{PATH}, library or Software Development Kit (SDK) version incompatibilities, regressions introduced by dependency updates, incorrect import paths, module name collisions, platform-specific dependency limitations, or container images missing required components. 
            The most frequent dependency-related problems are: \textit{Missing Dependency} (18.4\%), \textit{Backward Incompatibility/Breaking Change} (16.3\%), \textit{Executable Missing/Not on PATH} (14.3\%), \textit{Version Pin Used as Regression Workaround} (12.2\%), and \textit{Version Mismatch/Conflict} (8.2\%).

            \item \textit{Server Platform Incompatibility (21 issues)}: Issues where the MCP server fails or behaves inconsistently due to the OS, architecture, or other platform-specific differences in environment behavior, system utilities, encoding, or container support. This type of faults includes mismatches in OS-dependent path or encoding handling, missing platform-compatible packages or images, and integration problems. Among these, \textit{Microsoft Windows} is the most frequently affected platform (66.7\%), followed by \textit{Linux} (14.3\%) and \textit{macOS} (9.5\%).
            
        \end{enumerate}
    \item \textbf{Server Deployment (16 issues)}: Issues where the MCP server fails to build, package, install, or start due to problems in packaging metadata, build configuration, dependency bundling, distribution mechanisms, or deployment orchestration (e.g., Docker or package managers). These faults occur during the preparation or deployment of the server, before normal runtime execution, and include missing or mispackaged modules, incorrect build targets, distribution inconsistencies, or misconfigured container and environment settings.
    
    \item \textbf{Resource Management (10 issues)}: Issues related to the server’s access to and management of system
    resources, primarily files, directories, and shared project data, including concurrency issues and incorrect assumptions about resource availability or state. These issues are divided into two subgroups:
        \begin{enumerate}
            \item \textit{Server Access Synchronization (5 issues)}: Issues caused by concurrent or repeated access to shared resources, leading to race conditions, file locks, inconsistent states, or improper cleanup.
            
            \item \textit{Server Resource Handling (5 issues)}: Issues resulting from incorrect assumptions about the availability, structure, or state of resources such as files, directories, or workspaces, causing conflicts or inaccessible project data.
        \end{enumerate}
        
    \item \textbf{Server Logging (8 issues)}: Issues arising from misconfigured, malfunctioning, or intrusive server-side logging mechanisms that affect the observability or correct operation of the MCP server. These faults include incorrect or invalid log-level settings, logs emitted to unintended destinations, excessive or insufficient diagnostic output, logging-related exceptions (e.g., serialization or permission errors), or logging behavior that interferes with the server's JSON-over-stdout protocol communication.
    
    \item \textbf{Server Network Configuration (5 issues)}: Issues related to the server’s network communication setup or operation, including misconfigured HTTP request or response handling, header or metadata inconsistencies, or connection timeouts.
    
    \item \textbf{Application Programming Interface (API) Usage (5 issues)}: Issues caused by incorrect or invalid use of server-side APIs, including improper request or response construction, violations of API or schema contracts, missing or misconfigured parameters, or incorrect invocation of API-related commands or configurations. These issues arise when the MCP server interacts incorrectly with its own APIs or with external service
    APIs, leading to failures in request handling, model configuration, or API connectivity.
    
    \item \textbf{Server Infrastructure (1 issue)}: Issues in which the MCP server exhibits incorrect behavior or degraded performance due to faults, instability, or degradation in the underlying hardware or infrastructure (local or cloud-based), rather than faults in the server’s code or configuration. These problems arise from external infrastructure conditions such as platform outages, degraded storage or compute performance, or other environmental failures outside the server’s control.
\end{enumerate}

\subsubsection{\textbf{Server/Tool Configuration (133 issues)}}
This category encompasses issues that arise from how the MCP server defines, integrates, and manages its tools, as well as how those tools interact with external services, clients, and execution environments. These faults occur when tools are invoked, registered, authenticated, authorized, connected, or surfaced incorrectly, or when their dependencies, configurations, or responses do not align with the expectations of the MCP server or the host. Collectively, these issues reflect the challenges of maintaining consistent, secure, and interoperable tool behavior across diverse environments, protocols, and client implementations.
\begin{enumerate}
    \item \textbf{Tool Call/Execution (63 issues)}: This category includes issues that arise when invoking MCP tools, executing their logic, or coordinating interactions between tools, servers, and clients. These problems stem from misconfigured API or client endpoints, mismatches in parameter schemas or interface definitions, performance or concurrency limitations, incorrect execution logic, file system or path handling faults, improper SQL query or transaction management, platform-specific execution differences, or missing tool resources.
    The ten subgroups below capture the major sources of failures in tool invocation and execution workflows.
    \begin{enumerate}
        \item \textit{API/Endpoint \& Client Configuration (17 issues)}: Issues where server tools fail to communicate correctly with external APIs or client applications due to misconfigured endpoints, incorrect request or response handling, or protocol and version mismatches. These faults arise from missing or malformed parameters, invalid routes or URLs, unsupported request formats, or improper client-side configuration. The most prominent themes are: \textit{API Call Parameters} (17.6\%), \textit{API Response Handling} (17.6\%), \textit{URL/URI Formatting} (17.6\%), \textit{Connection Configuration} (11.8\%), and \textit{Query Formatting} (11.8\%).
        \item \textit{Parameter Schema/Type Validation (14 issues)}: Issues arising from inconsistencies between a tool’s declared parameter schema and the actual inputs it receives, including incorrect data types, malformed parameter structures, missing required fields, or misinterpreted optional arguments. These faults typically stem from mismatched schema definitions, incomplete or incorrect JSON Schema generation, or non-standardized serialization of parameters across tools.
        \item \textit{Stability, Concurrency, \& Performance (8 issues)}: Issues that affect the responsiveness or reliability of MCP servers and tools, and may manifest as CPU hangs, stalled or slow operations, connection reuse problems,
        and timeouts under heavy or concurrent workloads. These faults are primarily attributed to \textit{Scalability Limitations} (37.5\%), \textit{Algorithmic Non-Convergence} (25\%), \textit{Concurrency Handling} (25\%), and \textit{Memory Management} (12.5\%).
        \item \textit{Tool Interface \& Argument Parsing (7 issues)}: Issues arising from mismatched or outdated tool interfaces, including incorrect function signatures, inconsistent parameter parsing, and discrepancies between expected and actual argument structures across MCP tools, FastAPI endpoints, and Command-Line Interface (CLI) wrappers.
        \item \textit{Tool Execution Logic (5 issues)}: Issues arising from incorrect or incomplete execution logic within MCP tools, where an operation is logged as successful but does not produce the intended effect. These faults include ignored or improperly handled parameters, incorrect assumptions about the availability or state of required components, and issues in updating or applying changes to the underlying system state. Such issues stem from gaps in command implementation, state management, or capability checks within the tool.
        \item \textit{File System \& Path Handling (5 issues)}: Issues caused by incorrect or inconsistent handling of file paths, directories, or filesystem metadata within MCP tools. These include missing or improperly created directories, incorrect resolution of symlinks, invalid or overly restrictive path validation, and misinterpreted environment-based paths. Such faults arise from gaps in path normalization, directory management, or environment-aware file handling.
        \item \textit{Database Querying \& Transactions (3 issues)}: Issues arising from incorrect SQL query construction or transaction handling within MCP servers. These include improperly formatted or unsupported queries (66.67\%) and overly restrictive validation rules that reject valid SQL operations (33.33\%). Such faults typically stem from gaps in query parsing, transaction coordination, or schema-aligned validation.
        \item  \textit{Tests \& Example Scripts (2 issues)}: Issues arising from missing, outdated, or incomplete test coverage and example scripts, which lead to unvalidated tool behavior, integration gaps, or user confusion. These faults stem from insufficient test scaffolding or examples that do not reflect actual tool functionality.
        \item \textit{Platform-Specific Execution (1 issue)}: Issues caused by OS–specific behaviors, such as differences in line endings, file encodings, or system utilities, that lead to inconsistent tool execution or output across platforms.
        \item \textit{Missing Tool/Resource (1 issue)}: Issues caused by absent static files or other required resources in MCP tool distributions, leading to runtime errors or failed operations when these assets are accessed.
    \end{enumerate}
    
    \item \textbf{Tool Discovery/Registration (19 issues)}: Issues where tools defined in the MCP server are not made available in a usable form—either because registration is incomplete or incorrect, or because discovery and exposure mechanisms fail to surface the tool to clients. These faults cause tools to exist in code but remain undiscoverable, duplicated, missing, or inaccessible in practice. 
    \begin{enumerate}
        \item \textit{Tool Registration (16 issues)}: Issues where tools are incorrectly declared or registered on the server side, including faults in tool naming, schema structure, parameter definitions, or the timing and sequence of registration. In these cases, the tool exists in the codebase but is not exposed or recognized in a valid, discoverable, or usable form by the MCP server or host.
        \item \textit{Tool Exposure/Discovery (3 issues)}: Issues where tools are properly registered on the server but are not visible, discoverable, or listed by clients due to missing or incorrect capability declarations, naming or identification mismatches, or client-side filtering and interpretation rules that prevent the tools from being surfaced.
    \end{enumerate}
    
    \item \textbf{Tool Response Handling (18 issues)}: This category covers issues that arise when processing, validating, or interpreting the outputs returned by tools before they are forwarded to the host. These faults stem from unbounded or improperly segmented responses, schema/format mismatches, 
    poor error signaling and diagnostic feedback,
    inadequate exception handling, and unsafe prompt/input handling that introduces security risks. The four subgroups below capture the primary classes of tool response-related faults.
    \begin{enumerate}
        \item \textit{Response Size and Continuation Control (7 issues)}: Issues caused by tool responses that are excessively large, unbounded, or not properly segmented, including oversized JSON payloads, high-volume data objects, or long text outputs. These responses can exceed model context limits, trigger repeated continuation requests, or lead to degraded performance or stalls. Such faults arise from insufficient control over response size, streaming, or continuation boundaries.
        \item \textit{Schema and Format Compatibility Management (7 issues)}: Issues arising from mismatches in data structure, schema, or content format between tools and servers, including incorrect field names, incompatible shapes or layouts, or unsupported data types. These faults lead to parsing failures, validation errors, or improper interpretation of tool responses.
        \item \textit{Error Reporting and Robustness (3 issues)}: Issues where tools or servers fail to surface errors clearly or handle exceptional conditions safely, including missing or undefined methods, uninformative messages, and unhandled exceptions. These faults lead to crashes or user confusion due to inadequate diagnostic feedback or insufficient defensive handling.
        \item \textit{Security and Prompt Sanitization (1 issue)}: Issues involving unsafe or unsanitized handling of user inputs in prompts or tool configurations, including improper variable interpolation, unvalidated parameters, or prompt structures that expose tools to injection or other security risks. These faults arise when input handling fails to enforce safety constraints or isolate untrusted content.
    \end{enumerate}

    \item \textbf{Tool Authorization (11 issues)}: Issues where tools lack the required permissions to access resources or perform specific actions, or where authorization policies are incorrectly applied or propagated. These faults include missing or insufficient privileges, misinterpreted access rules, improper handling or forwarding of authorization credentials, and tools performing actions outside their permitted scope. Such issues lead to authorization failures, blocked operations, or unintended access to protected resources.

    \item \textbf{Tool Authentication (8 issues)}: Issues arising from faults in establishing or validating the credentials required for tools to authenticate with external services or identity providers. These faults include incorrect or missing authentication parameters, misconfigured or outdated credential settings, improper handling or propagation of authentication headers or tokens, and mismatches between expected and actual authentication flows. Such issues lead to failed logins, blocked API access, or disrupted authentication handshakes during tool invocation.
    
    \item \textbf{Tool Dependency (8 issues)}: Issues caused by missing, outdated, or incompatible dependencies required specifically for tool functionality, distinct from the MCP server’s own dependencies. These faults arise when tools rely on external APIs, libraries, runtimes, or platform features that are unavailable, deprecated, or mismatched in version, leading to runtime errors, attribute or method failures, or tools not loading or executing as expected.
    
    \item \textbf{Tool Connection Configuration (6 issues)}: Issues stemming from incorrect or unstable configuration of the connection between the MCP server and the resources invoked by server-defined tools. These faults arise when the server cannot reliably initiate, maintain, or interpret the communication channel due to misconfigured connection parameters, incompatible environment settings, or timing- and load-related connection failures. This category consists of two subgroups:
        \begin{enumerate}
            \item \textit{Incorrect Tool Connection Setting (5 issues)}: Issues caused by misconfigured parameters or environment settings required to establish a valid connection between a server-defined tool and the resource, service, or API it invokes. These include incorrect command paths, incompatible or outdated integration scripts, missing runtime dependencies, improper handling of connection strings, or conflicts arising from multiple active tool instances or ports. Such faults prevent the server from initializing or sustaining the expected communication channel.
            \item \textit{Tool Connection Synchronization (1 issue)}: Issues arising from incorrect timing or coordination of tool connections, including failures under concurrent load, delayed responses, or mismatches between expected and actual connection lifecycles. These faults occur when a server-defined tool and the invoked resource fall out of synchronization during connection establishment or execution, leading to premature termination or unstable communication.
        \end{enumerate}
\end{enumerate}

\subsubsection{\textbf{Server/Host Configuration (120 issues)}} 
\label{sec:server_host}
This category includes issues arising from how MCP hosts (IDEs, apps, UI shells) are configured to connect to, launch, and interact with MCP servers. These problems occur when connection settings, configuration files, environment variables, dependency versions, transport modes, or LLM integrations are misaligned between host and server—leading to failed startups, unstable connections, misrouted requests, or incorrect interpretation of server responses. The main subcategories are:
\begin{enumerate}
    \item \textbf{Host Connection Configuration (45 issues)}: Issues involving the establishment, maintenance, and termination of the communication channel between the MCP server and the host. These faults arise when connection protocols, configuration parameters, timing expectations, or session-management mechanisms do not align between the two sides, leading to unstable, delayed, or inconsistent host–server communication. The following subcategories capture the major forms of connection-related failures. 
        \begin{enumerate}
            \item \textit{Host Connection Synchronization (16 issues)}: Issues caused by timing, ordering, or life-cycle mismatches in the host–server communication flow. These faults occur when requests are sent before the server has fully initialized, when long-running tasks block subsequent communication, when multiple concurrent connections interfere with each other, or when connection reuse or premature process termination disrupts message exchange. Such problems manifest as hangs, timeouts, race conditions, or stalled connections during initialization or request processing.
            
            \item \textit{Host Connection Configuration Mismatch (13 issues)}: Issues caused by incompatibilities between the host’s expected communication protocol, transport mode, or message format and the MCP server’s actual implementation. These faults arise when the host and server disagree on connection type, endpoint usage, HTTP method, request or response structure, or the formatting of JSON-RPC messages, leading to failed initialization, malformed message exchanges, or inability to establish a stable connection.
            
            \item \textit{Incorrect Host Connection Setting (13 issues)}: Issues caused by misconfigured or incomplete parameters required to establish or maintain the host–server connection. These faults arise when the host or MCP server uses incorrect URLs, routes, environment variables, configuration fields, or startup procedures; when required connection settings are missing or misnamed; or when
            MCP clients construct host–server endpoints improperly. Such misconfigurations lead to failed handshakes, rejected initialization responses, unexpected exceptions during connection setup, or timeouts when the server cannot be reached or launched correctly from the host environment.

            \item \textit{Session Handling (3 issues)}: Issues related to the management of multiple host sessions, including tracking, isolating, persisting, or cleaning up session state. These faults lead to inconsistent behavior when multiple MCP clients interact with the server concurrently or sequentially.

                \begin{itemize}
                    \item \textit{Session Isolation (2 issues)}: Issues where session data or contextual information leaks between concurrent or successive MCP client sessions due to insufficient separation of workspace, project, or state-specific information. These faults cause actions or outputs in one session to reference or affect the state of another, indicating incomplete isolation mechanisms.
                    \item \textit{Session Update (1 issue)}: Issues arising when session state is not properly refreshed or synchronized after external configuration changes, such as switching workspaces or updating active projects. These faults lead to stale or inconsistent session information being used during tool execution or host–server interactions.
                \end{itemize}
        \end{enumerate}
        
    \item \textbf{Host's Server Configuration (37 issues)}: 
    Issues where the MCP host misconfigures, launches incorrectly, or provides incorrect settings to the MCP server through host-managed configuration files, environment variables, executable paths, arguments, or transport definitions. These faults arise when the server inherits an invalid configuration state from the host, causing server startup failures, initialization errors, or unexpected connection behavior. The main types of issues identified within this subcategory are:
    \begin{enumerate}
        \item \textit{Executable Path Resolution (10 issues)}: Issues where the MCP host cannot launch or locate the MCP server because required executables, command paths, or environment-resolved binaries are missing, incorrect, or inaccessible. These faults arise when the server configuration points to non-existent or relative paths, when executables are not included in the host’s environment (e.g., PATH, virtual environments, WSL/Docker namespaces), or when dependencies cannot be spawned due to filesystem or environment constraints.
        \item \textit{URL/Connection String Formatting (7 issues)}: Issues caused by incorrectly structured or incomplete URLs, URIs, or file paths in MCP host configuration. These faults include wrong or missing prefixes or suffixes, invalid protocol or SSL settings, and misformatted connection strings that prevent the host from resolving the server endpoint or accessing required resources.
        \item \textit{Environment Variable Validation (5 issues)}: Issues where MCP servers fail to start, initialize, or connect because required environment variables configured through the host
        are missing, misnamed, incorrectly formatted, or interpreted too strictly. These faults arise from invalid variable values (e.g., unsupported characters or spacing), variables marked as required despite being optional, incorrect working-directory assumptions requiring explicit path variables, or misconfigured connection-related or region-related settings that rely on environment-based configuration.
        \item \textit{Command/Argument Semantics (5 issues)}: Issues arising from incorrect, ambiguous, or platform-dependent interpretation of command-line syntax, arguments, or flags used
        by MCP host to start or configure MCP servers. These include improperly structured flags, unsupported argument combinations, shell-specific parsing differences, and incorrect assumptions about available CLI commands. Such faults lead to commands being rejected, misinterpreted, or unable to locate the intended executables or configurations.
        \item \textit{Configuration File Location/Selection (4 issues)}: Issues caused by MCP configuration or environment files being placed in the wrong directory or given incorrect names in the MCP host.
        These faults lead MCP servers or clients to load the wrong settings, ignore required variables, or fail during startup or validation because the expected configuration source is not found. Such issues arise when tools expect configuration files in specific global locations, impose strict schema rules, or rely on well-defined file naming and placement conventions.
        \item \textit{Transport Mode Misconfiguration (3 issues)}: Issues caused by incorrect or incomplete configuration of the communication transport between the MCP host and server, such as specifying the wrong mode (e.g., SSE vs. STDIO), omitting required fields, or providing incompatible endpoint definitions.
        These faults lead to failures in establishing or maintaining the connection, incorrect routing of messages, or improper handling of requests due to the MCP server and client operating under mismatched transport expectations.
        \item \textit{JSON Syntax/Structure Errors (2 issues)}: Issues caused by malformed, incomplete, or improperly saved JSON configuration files, such as trailing commas, invalid formatting, or unsynchronized configuration state. These faults prevent the MCP host from parsing MCP configuration files correctly, leading to initialization failures or rejection of the configuration during server startup.
        \item \textit{Host-side Overwrite/Merge Behavior (1 issue)}: Issues where the MCP host unintentionally overwrites, resets, or merges configuration files or environment settings during server startup or initialization.
        Such behavior replaces user-defined values with defaults or recomputed settings, leading to incorrect configuration states or server startup
        failures.
    \end{enumerate}
    
    \item \label{subsec:LLM}
    \textbf{LLM (17 issues)}: Issues originating from the behavior of, or integration with, the underlying LLM used by the MCP host. These include limitations in prompt interpretation, incorrect or inconsistent function-call generation, hallucinated commands or content, integration failures with external LLM providers, and problems caused by exceeding model context limits. The main types of issues in this category are:
    \begin{enumerate}
        \item \textit{Model Capability and Prompt Interpretation (5 issues)}: Issues arising from limitations in the language model’s ability to follow instructions, reason over complex prompts, or preserve input fidelity. This subcategory includes misinterpreted or altered inputs, incorrect or unstable outputs, or failures to adhere to required formats or task structures. Such faults occur when the model lacks the capability, precision, or consistency needed for reliable tool-driven operations or structured-response workflows.
        \item \textit{Function Invocation and Parameter Handling (4 issues)}: Issues where the language model generates incorrect, incomplete, or outdated function calls or parameter structures. These faults include malformed or misformatted inputs, incorrect or mismatched function names, invalid argument types, or code that does not align with the tool’s current API or schema. Such errors occur when the LLM improperly maps natural-language instructions to formal tool calls, leading to execution failures or protocol-level errors.
        \item \textit{Hallucinated Command/Content (3 issues)}: Issues where the language model fabricates commands, file paths, URLs, or other external references that do not exist in the actual environment. These faults occur when the LLM substitutes imagined or training-data–derived artifacts for real system resources, leading to invalid tool calls, inaccessible references, or inconsistencies between model output and the host filesystem or API context.
        \item \textit{LLM Provider Integration Failure (3 issues)}: 
        Issues where the MCP host fails to correctly integrate with an external LLM provider due to service unavailability, incorrect API or endpoint configuration, or mismatched provider-specific settings. These faults occur when the server cannot successfully route requests to the intended LLM backend, resulting in incomplete outputs, request errors, or degraded tool behavior.
        \item \textit{Context Length Limit (2 issues)}: Issues where the language model reaches or exceeds its maximum context window during large input processing or extended generation, leading to incomplete outputs, repeated restarts, or stalled execution. These faults arise when the size of the content, conversation state, or generated output surpasses the model’s effective context capacity, preventing reliable continuation or completion of long operations.
    \end{enumerate}
    
    \item \textbf{Host Dependency (11 issues)}: Issues where the MCP host fails to start, load, or communicate with an MCP server due to missing, incompatible, or incorrectly installed dependencies in the environment from which the host launches the MCP server. These faults arise when host-side updates introduce breaking changes, when required modules or executables are absent or mispackaged, or when server dependencies fail at import or initialization time during host-driven startup. Such issues typically manifest as import errors, schema-validation failures,
    unexpected exceptions, or server startup failures triggered by dependency mismatches between the host environment and the MCP server.
    
    \item \textbf{Host Logging (8 issues)}: Issues arising from incorrect handling of server-side log output by the MCP host, particularly when log messages interfere with the JSON-RPC communication stream. These faults occur when MCP servers emit logs to \texttt{stdout} instead of \texttt{stderr}, use overly verbose or incompatible log levels, or generate non-JSON diagnostic output that the host attempts to parse as protocol messages. Such issues lead to JSON parsing failures, disrupted initialization, or inconsistent host behavior when interpreting MCP server responses.
    
    \item \textbf{Host's Hook Configuration (1 issue)}: Issues caused by incorrect configuration of lifecycle hooks or event-driven actions defined on the MCP host. These faults occur when user-configured shell commands or automation scripts reference invalid paths, use machine-specific settings, or are otherwise misaligned with the host environment, leading to failed hook execution or broken host behavior during key lifecycle events.
    
    \item \textbf{Host Authorization (1 issue)}: Issues where the MCP server cannot access required resources or perform host-mediated operations because the underlying platform denies permission. These faults arise when the service account, process, or identity under which the server runs lacks the necessary privileges for catalog, workspace, or resource-level actions, leading to startup failures or blocked requests until proper host-level permissions are granted.
    
\end{enumerate}

\subsubsection{\textbf{Documentation (29 issues)}}
This category includes issues raised in the formal documentation provided for MCP servers, such as \texttt{README} files, as well as issues related to tutorials or example codes or configurations of MCP servers.

\subsubsection{\textbf{General Programming (22 issues)}} This category includes issues consisting of typical software development errors not specific to the MCP architecture (e.g., null-pointer crashes, typos, syntax errors), as well as user mistakes in writing or configuring MCP servers.

\begin{keyfindingsbox}{RQ1 - Key Findings}
\begin{itemize}
    \item The final taxonomy contains five high-level categories and 419 manually assigned labels, with some issues receiving multiple labels when they involved more than one MCP-related fault.
    \item Most MCP faults fall into three boundary-oriented categories: \textit{Server/Tool Configuration} (31.74\%), \textit{Server/Host Configuration} (28.64\%), and \textit{Server Setting} (27.45\%).
    \item \textit{Documentation} (6.92\%) and \textit{General Programming} (5.25\%) are less frequent, indicating that most observed faults are tied to MCP-specific configuration and interaction concerns rather than ordinary implementation mistakes alone.
    \item The taxonomy shows that MCP faults frequently involve mismatches among server behavior, host expectations, tool schemas, transport choices, dependencies, and execution environments.
\end{itemize}
\end{keyfindingsbox}

\subsection{Taxonomy Validation (RQ2)}

To validate the proposed taxonomy of faults in MCP servers, we conduct a survey among MCP practitioners. In total, 41 valid responses were collected. The participants’ roles included developers (46.51\%), researchers (13.95\%; e.g., postgraduate students, professors, and research associates), development team managers (18.60\%), software engineers (16.28\%), and data scientists (4.65\%). Regarding implementation technologies, participants reported that Python (42.47\%), TypeScript (23.29\%), and JavaScript (9.59\%) are the most commonly used programming languages for developing MCP servers. We interpret the survey as complementary evidence to the GitHub issue analysis: the issue corpus captures faults that become public development issues, whereas the survey captures practitioners' direct exposure to the taxonomy categories and their perceived severity and remediation effort.

\begin{table*}[t]
\caption{Results of the survey of MCP practitioners.}
\label{tbl:validation_report}
\centering
\resizebox{0.98\textwidth}{!}{
\begin{tabular}{c | p{4.8cm} | c | ccc | ccc | cc}
\toprule
\multirow{2}{*}{\textbf{Category}} &
\multirow{2}{*}{\textbf{Fault Type}} & 
\multicolumn{1}{c|}{\textbf{Prevalence}} &
\multicolumn{3}{c|}{\textbf{Severity}} &
\multicolumn{3}{c|}{\textbf{Required Effort}} &
\multicolumn{2}{c}{\textbf{Index}} \\
\cmidrule(lr){3-3}
\cmidrule(lr){4-6} \cmidrule(lr){7-9} \cmidrule(lr){10-11}
 &  & Response Rate & Minor & Major & Critical & Low & Medium & High & Sev. & Eff. \\
\midrule
\multirow{9}{*}{\rotatebox{90}{Server Setting}} 
& Server Deployment & 53.66\%  & 48.15\% & 33.33\% & 18.52\% & 41.38\% & 37.93\% & 20.69\% & 0.35 & 0.40 \\
& Server Network Configuration & 48.78\% & 64.29\% & 25.00\% & 10.71\% & 70.00\% & 16.67\% & 13.33\% & 0.23 & 0.22 \\
& API Usage & 60.98\% & 51.72\% & 31.03\% & 17.24\% & 44.83\% & 34.48\% & 20.69\% & 0.33 & 0.38 \\
& Server Dependency & 43.90\% & 59.26\% & 25.93\% & 14.81\% & 62.96\% & 22.22\% & 14.81\% & 0.28 & 0.26 \\
& Server Platform Incompatibility & 48.78\%  & 53.85\% & 15.38\% & \cellcolor{blue!10}\textbf{30.77\%} & 51.85\% & 25.93\% & 22.22\% & 0.38 & 0.35 \\
& Server Infrastructure & 29.27\%  & 62.50\% & 12.50\% & \cellcolor{blue!10}\textbf{25.00\%} & 64.00\% & 8.00\% & \cellcolor{blue!10}\textbf{28.00\%} & 0.31 & 0.32 \\
& Server Logging & 39.02\% & 72.00\% & 16.00\% & 12.00\% & 46.15\% & 42.31\% & 11.54\% & 0.20 & 0.33 \\
& Server Resource Synchronization & 19.51\% & 71.43\% & 19.05\% & 9.52\% & 72.73\% & 18.19\% & 9.09\% & 0.19 & 0.18 \\
& Server Resource Handling & 41.46\% & 40.91\% & 45.45\% & 13.64\% & 56.52\% & 30.43\% & 13.04\% & 0.36 & 0.28 \\
\midrule
\multirow{8}{*}{\rotatebox{90}{\parbox{2.2cm}{\centering Server / Tool Configuration}}}
& Tool Authentication & 38.24\% & 56.52\% & 21.74\% & 21.74\% & 52.17\% & 13.04\% & \cellcolor{blue!10}\textbf{34.78\%} & 0.33 & 0.41 \\
& Tool Authorization & 20.59\% & 55.00\% & 20.00\% & \cellcolor{blue!10}\textbf{25.00\%} & 55.00\% & 20.00\% & 25.00\% & 0.35 & 0.35 \\
& Tool Call / Execution & 52.94\% & 56.52\% & 30.43\% & 13.04\% & 52.17\% & 34.78\% & 13.04\% & 0.28 & 0.30 \\
& Tool Response Handling & \cellcolor{blue!10}\textbf{66.67\%}& 47.83\% & 34.78\% & 17.39\% & 43.48\% & 34.78\% & 21.74\% & 0.35 & 0.39 \\
& Tool Discovery \& Registration & 27.27\% & 57.89\% & 10.53\% & \cellcolor{blue!10}\textbf{31.58\%} & 38.89\% & 50.00\% & 11.11\% & 0.37 & 0.36 \\
& Tool Dependency & 24.24\% & 61.11\% & 16.67\% & 22.22\% & 61.11\% & 16.67\% & 22.22\% & 0.31 & 0.31 \\
& Tool Connection Synchronization & 39.39\% & 55.00\% & 35.00\% & 10.00\% & 60.00\% & 25.00\% & 15.00\% & 0.28 & 0.28 \\
& Incorrect Tool Connection Setting & 24.24\% & 68.42\% & 21.05\% & 10.53\% & 63.16\% & 26.32\% & 10.53\% & 0.21 & 0.24 \\
\midrule
\multirow{11}{*}{\rotatebox{90}{Server/Host Configuration}}
& Host's Server Config & 53.33\% & 63.16\% & 26.32\% & 10.53\% & 73.68\% & 15.79\% & 10.53\% & 0.24 & 0.18 \\
& Host Logging & \cellcolor{blue!10}\textbf{13.79\%}  & 86.67\% & 6.67\% & 6.67\% & 73.33\% & 26.67\% & \cellcolor{blue!10}\textbf{0.00\%} & 0.10 & 0.13 \\
& Large Language Model (LLM) & 53.33\% & 56.52\% & 34.78\% & 8.70\% & 50.00\% & 27.27\% & 22.73\% & 0.26 & 0.36 \\
& Host Dependency & 40.00\% & 55.56\% & 38.89\% & \cellcolor{blue!10}\textbf{5.56\%} & 61.11\% & 27.78\% & 11.11\% & 0.25 & 0.25 \\
& Host's Hook Config & \cellcolor{blue!10}\textbf{16.67\%} & 85.71\% & 14.29\% & \cellcolor{blue!10}\textbf{0.00\%} & 85.71\% & 14.29\% & \cellcolor{blue!10}\textbf{0.00\%} & 0.07 & 0.07 \\
& Host Authorization & \cellcolor{blue!10}\textbf{13.33\%} & 80.00\% & 20.00\% & \cellcolor{blue!10} \textbf{0.00\%} & 78.57\% & 7.14\% & 14.29\% & 0.10 & 0.18 \\
& Host Connection Mismatch & 43.33\% & 38.89\% & 38.89\% & 22.22\% & 61.11\% & 27.78\% & 11.11\% & 0.42 & 0.25 \\
& Incorrect Host Connection Setting & 23.33\% & 46.67\% & 40.00\% & 13.33\% & 66.67\% & 26.67\% & 6.67\% & 0.33 & 0.20 \\
& Host Connection Synchronization & 36.67\% & 53.33\% & 40.00\% & 6.67\% & 53.33\% & 40.00\% & 6.67\% & 0.27 & 0.27 \\
& Session Isolation & 20.00\% & 60.00\% & 26.67\% & 13.33\% & 66.67\% & 20.00\% & 13.33\% & 0.27 & 0.23 \\
& Session Update & 23.33\% & 53.33\% & 40.00\% & 6.67\% & 53.33\% & 46.67\% & \cellcolor{blue!10}\textbf{0.00\%} & 0.27 & 0.23 \\
\midrule
Doc & Documentation & \cellcolor{blue!10}\textbf{64.52\%} & 58.33\% & 33.33\% & 8.33\% & 54.17\% & 33.33\% & 12.50\% & 0.25 & 0.29 \\
\bottomrule
\end{tabular}
}
\vspace{0.35ex}
\noindent\begin{minipage}{0.98\textwidth}
\footnotesize\linespread{0.95}\selectfont\raggedright
\textit{Note.} Severity and effort indices are normalized weighted summaries of ordered survey responses~\cite{holt2014asking,perera2022factors}. We compute $\smash{\text{Sev. idx.}=(\text{Major}+2\times\text{Critical})/2}$ and $\smash{\text{Eff. idx.}=(\text{Medium}+2\times\text{High})/2}$. Values range from 0 to 1, with larger values indicating higher perceived severity or effort.
\end{minipage}
\end{table*}

\subsubsection{Taxonomy Coverage and Practitioner Prevalence}
Table~\ref{tbl:validation_report} summarizes survey participants’ responses for each MCP fault type identified in the proposed taxonomy. \textbf{Notably, all MCP-specific fault types defined in the taxonomy are encountered by at least some MCP practitioners}, providing empirical support for the completeness of the taxonomy. Among the identified fault types, \textit{tool response handling} (66.67\%) and \textit{documentation} (64.52\%) are reported by participants as the most frequent fault types. Several other fault types also have high reported prevalence, including \textit{API usage} (60.98\%), \textit{server deployment} (53.66\%), \textit{host's server configuration} (53.33\%), \textit{Large Language Model (LLM)} faults (53.33\%), and \textit{tool call/execution} (52.94\%). This pattern suggests that practitioners do not experience MCP faults as a single dominant failure mode; rather, recurring problems appear across server setup, tool interaction, host-side configuration, and documentation.

The relationship between survey prevalence and taxonomy frequency is also informative. Because the issue-derived taxonomy occurrence percentages did not satisfy the normality assumption according to the Shapiro--Wilk test~\cite{shapiro1965analysis}, we use Spearman rank correlation to compare taxonomy occurrence with practitioner-reported prevalence across the 29 fault types shared by the taxonomy and survey. The two measures show a strong positive association ($\rho=0.587$, $p=0.001$) under conventional correlation thresholds, indicating that fault types appearing more often in the GitHub-derived taxonomy also tend to be encountered more often by practitioners. At the same time, examples of divergence show that the two sources are not redundant. For example, \textit{tool call/execution} and \textit{tool discovery \& registration} appear frequently in the taxonomy, whereas practitioners report \textit{tool response handling} as more prevalent. This difference is expected because issue counts reflect faults that are reported and resolved in public repositories, whereas survey responses reflect practitioners' direct exposure across development and usage settings. Tool response handling may be especially visible to practitioners because it occurs after a tool has already been invoked: the failure may appear as malformed, oversized, incomplete, or semantically unusable output returned through the MCP server to the host. Thus, even when tool registration and invocation succeed, the end-to-end interaction can still fail at the response boundary. This finding suggests that MCP testing should cover not only tool discovery and call execution, but also response contracts, output-size limits, schema conformance, error representation, and host-side parsing of returned tool results.

Furthermore, the high prevalence of documentation-related faults may be attributed to the relative novelty of MCP, as its supporting documentation is still evolving. Consistent with the taxonomy results, documentation ranks as the fourth most frequent MCP-related fault category overall. At the same time, documentation has comparatively low reported criticality (8.33\% critical) and high-effort remediation (12.50\% high effort), indicating that documentation faults are common usability barriers but are not generally perceived as the most severe or difficult faults to repair.

In contrast, \textit{host authorization} (13.33\%), \textit{host logging} (13.79\%), and \textit{host hook configuration} (16.67\%) are reported as the least frequent fault types by survey participants. These findings are aligned with the taxonomy analysis, which similarly identifies \textit{host hook configuration} and \textit{host authorization} as the least prevalent MCP-related faults.

\subsubsection{Severity and Required Effort}
Beyond whether participants had encountered each fault type, the survey also asked them to assess perceived severity (minor, major, or critical) and required remediation effort (low, medium, or high). To compare these ordered responses across fault types, Table~\ref{tbl:validation_report} reports both the full response distributions and two normalized indices, following prior uses of weighted indices for ordered survey responses~\cite{holt2014asking,perera2022factors}: the severity index, computed as $(\textit{Major}+2\times\textit{Critical})/2$, and the effort index, computed as $(\textit{Medium}+2\times\textit{High})/2$. Both indices range from 0 to 1, with larger values indicating higher perceived severity or effort. Using the severity index, \textit{host connection mismatch} has the highest perceived severity (0.42), showing that severity provides a different view of MCP faults than prevalence alone. This result is consistent with the role of the host as the component that connects to, configures, and invokes MCP servers: when the host and server disagree about connection settings or expected interaction behavior, the resulting failure can prevent reliable communication across the MCP boundary rather than remaining a localized configuration issue.

The next highest severity-index values are observed for \textit{server platform incompatibility} (0.38) and \textit{tool discovery \& registration} (0.37). These two fault types are also important under a stricter critical-only view: \textit{tool discovery \& registration} has the highest critical-response share (31.58\%), followed closely by \textit{server platform incompatibility} (30.77\%). This indicates that the severity index and the critical-only signal emphasize related but not identical risks. \textit{Tool discovery \& registration} is especially concerning in worst-case scenarios because, if the host cannot identify or register the tools exposed by a server, later stages of the interaction cannot proceed. \textit{Server platform incompatibility}, in turn, combines relatively high prevalence (48.78\%) with high criticality, suggesting that platform assumptions remain a practical risk for MCP servers across runtime, operating-system, dependency, and deployment settings.

At the lower end of the severity index, \textit{host's hook configuration} (0.07), \textit{host logging} (0.10), and \textit{host authorization} (0.10) are perceived as lower-impact faults. Overall, the severity analysis refines the prevalence-based picture: common faults such as \textit{tool response handling} and \textit{documentation} matter because many practitioners encounter them, whereas faults such as \textit{host connection mismatch}, \textit{tool discovery \& registration}, and \textit{server platform incompatibility} deserve particular attention because they can interrupt core MCP functionality or prevent servers from operating reliably across target environments.

The required-effort responses show a related but distinct prioritization pattern. Considering the effort index in Table~\ref{tbl:validation_report}, \textit{tool authentication} has the highest perceived remediation effort (0.41), followed closely by \textit{server deployment} (0.40), \textit{tool response handling} (0.39), and \textit{API usage} (0.38). These high-index fault types point to work that crosses component or environment boundaries: authentication faults require agreement on credentials, handshake behavior, and security configuration; deployment and API-usage faults depend on the runtime and integration context; and response-handling faults require validating the structure, size, and semantics of tool outputs returned to the host.

The high-effort response share provides additional nuance. \textit{Tool authentication} also has the largest share of high-effort responses (34.78\%), reinforcing its position as the most effort-intensive fault type. \textit{Server infrastructure} has a notable high-effort tail as well (28.00\%), even though its overall effort index is lower (0.32), suggesting that infrastructure faults may be polarized: many are handled with relatively low effort, but some require more extensive environment or deployment-level debugging.

At the lower end of the effort index, \textit{host's hook configuration} (0.07), \textit{host logging} (0.13), \textit{host's server configuration} (0.18), and \textit{host authorization} (0.18) require comparatively less remediation effort. These faults are often addressable through localized host-side changes, such as revising configuration files, adjusting access rules, or improving logging settings, rather than through broader protocol, deployment, or cross-component revisions. Overall, the effort analysis suggests that the most time-consuming MCP faults are not simply the most severe ones; they are often the faults that require coordinating behavior across authentication, deployment, API, and response boundaries.

\subsubsection{Cross-Signal Statistical Checks}
To further examine whether the survey dimensions capture overlapping or distinct aspects of MCP faults, we computed Spearman correlations across the 29 fault types using the prevalence response rate and these severity and effort indices. Severity and required effort are positively correlated ($\rho=0.65$, $p<0.001$), indicating that fault types perceived as more severe also tend to require more remediation effort. However, the relationship is not one-to-one. For example, \textit{host connection mismatch} has the highest severity index (0.42) but only a moderate effort index (0.25), whereas \textit{tool authentication} has the highest effort index (0.41) while ranking lower in severity (0.33). This suggests that remediation effort is shaped not only by perceived impact, but also by debugging complexity, credential/configuration dependencies, and environment-specific reproduction barriers. Prevalence shows a weaker positive association with severity ($\rho=0.37$, $p=0.049$) and a stronger association with required effort ($\rho=0.57$, $p=0.002$). These correlations indicate a broad rank-level tendency, not that the most common faults are necessarily the most severe or effort-intensive. They should therefore be read together with the exceptions discussed above, where several frequent faults have modest severity or effort indices, and several less frequent faults remain important because of their criticality or remediation complexity.

We further compared the survey dimensions across the three high-level configuration groups, excluding \textit{Documentation} because it is not part of these groups.
Using the index values, we first applied the \textit{Shapiro--Wilk} test~\cite{shapiro1965analysis}; since the normality assumption was not rejected, we used one-way ANOVA~\cite{fisher1925statistical} followed by Tukey--Kramer post-hoc comparisons~\cite{tukey1949comparing,kramer1956extension}. The group differences were not statistically significant for prevalence ($p=0.203$) or severity index ($p=0.125$). In contrast, effort index differed significantly across groups ($p=0.004$), with a large effect size ($\eta^2=0.359$) according to conventional eta-squared thresholds~\cite{richardson2011eta,cohen_statistical_2013}. The post-hoc comparisons localized this difference to \textit{Server/Host Configuration}, which has lower effort indices than \textit{Server Setting} and \textit{Server/Tool Configuration}, while \textit{Server Setting} and \textit{Server/Tool Configuration} were not significantly different from each other. This pattern suggests that host-side faults may often be localized once diagnosed, even though the issue-discussion evidence in RQ3 shows that reproducing user-specific host environments can still require substantial diagnostic back-and-forth.

\subsubsection{Open-Ended Validation}
Several participants suggested additional fault types that, upon closer examination, are already encompassed by the proposed taxonomy. For instance, one participant noted that different LLM versions may exhibit distinct behaviors when interacting with external tools. This observation maps directly to the \textit{LLM} subcategory within the \textit{server/host configuration} category. As discussed in Subsection~\ref{sec:server_host}, this subcategory includes faults such as \textit{LLM Provider Integration Failure}, which captures inconsistencies, behavioral variations, and incompatibilities stemming from differences across LLM providers or model versions.
Similarly, another participant highlighted faults related to the use of MCP SDKs with differing performance characteristics. These issues fall under the \textit{server dependency} subcategory described in Subsection~\ref{sec:server_setting}. This subcategory explicitly includes faults stemming from dependencies on specific SDKs, libraries, or external components.

Overall, all faults reported by survey participants can be systematically mapped to existing categories within the proposed taxonomy. In addition, each fault type in the taxonomy was reported by at least a subset of practitioners, indicating that the categories are grounded in real-world experience. The absence of newly suggested and uncategorized fault types provides further evidence supporting the coverage and practical completeness of the proposed taxonomy while acknowledging that future ecosystem evolution may surface additional categories.

\begin{keyfindingsbox}{RQ2 - Key Findings}
\begin{itemize}
    \item All fault types in the taxonomy were encountered by at least some practitioners, supporting the taxonomy's practical coverage.
    \item Issue-derived taxonomy occurrence and practitioner-reported prevalence are strongly aligned, while still providing complementary views of MCP faults.
    \item \textit{Tool response handling} and \textit{documentation} are the most prevalent practitioner-facing faults, whereas \textit{host connection mismatch}, \textit{tool discovery \& registration}, and \textit{tool authentication} rank highest by severity, critical-response share, and remediation effort, respectively.
    \item At the high-level group level, required effort differs significantly across the three configuration groups, with \textit{Server/Host Configuration} showing lower effort indices than the other two groups.
\end{itemize}
\end{keyfindingsbox}

\subsection{Characteristics of MCP-related Faults (RQ3)}\label{subsec:fault_characteristics}

To analyze the characteristics of MCP-related faults, we employ a set of issue-centric metrics commonly used in prior empirical studies~\cite{kononenko2018studying,morovati2024bug}. These metrics include (1) required time to fix the issue, (2) number of comments, (3) number of collaborators involved in the discussion, (4) average number of comments per collaborator, and (5) the experience level of the developer who fixes the issue.
Required time to fix the issue is computed as the duration between the issue’s opening and closing dates, following established methodologies in the literature~\cite{bosu2014impact}. 
The number of collaborators for each closed issue is measured as the count of unique individuals who commented on or contributed to the discussion regarding potential solutions.
Consistent with previous work, developer experience (i.e., experience of fixer) is approximated using the number of commits made by the developer immediately prior to closing the issue~\cite{romano2021empirical,morovati2024bug}.

To identify whether the differences in these metrics across fault types are statistically significant, we first assess the distribution of our data to determine the appropriate statistical test. Using the \textit{Shapiro–Wilk} test~\cite{shapiro1965analysis}, we find that the data do not meet the normality assumption. Accordingly, we rely on non-parametric statistical methods.
To compare the characteristics of the four main MCP-related fault types plus \textit{General Programming} category, we apply the \textit{Kruskal–Wallis} test~\cite{kruskal_use_1952,mckight_kruskal-wallis_2010}, which evaluates differences among three or more independent groups. 
Given that relying solely on \textit{p-values} can be misleading~\cite{kampenes2007systematic}, we also compute the \textit{Eta-squared ($\eta^2$)} effect size~\cite{richardson2011eta} for the \textit{Kruskal–Wallis} test.
In general, effect size allows us to assess the magnitude and practical significance of the observed differences. 
\textit{Eta-squared} represents the proportion of variance explained by the grouping factor and is a recommended overall effect-size measure for rank-based tests because it provides an interpretable estimate on a bounded $[0,1]$ scale~\cite{tomczak_need_2014}.
\textit{Eta-squared} values of approximately 0.01, 0.06, and 0.14 are commonly interpreted as small, medium, and large effect sizes, respectively~\cite{cohen_statistical_2013}.
When the \textit{Kruskal–Wallis} test identifies statistically significant differences, we perform post-hoc pairwise comparisons using \textit{Dunn’s} test~\cite{dunn_multiple_1964} with \textit{Bonferroni} correction~\cite{bonferroni_when_2014}. 
Given the importance of reporting effect sizes, we compute the \textit{Vargha–Delaney A$_{12}$}
effect size~\cite{vargha2000critique} for all \textit{Dunn's} post-hoc comparisons to quantify the magnitude of the observed differences.
\textit{Vargha–Delaney A$_{12}$} represents the probability that a randomly selected observation from one group exceeds a randomly selected observation from another group, making it a natural complement to \textit{Dunn’s} rank-based post-hoc test. An \textit{A$_{12}$} value of 0.5 indicates no effect, while values in the ranges $0.56$–$0.64$ (or $0.36$–$0.44$) indicate small effects, $0.64$–$0.71$ (or $0.29$–$0.36$) indicate medium effects, and values above $0.71$ (or below $0.29$) reflect large effects~\cite{vargha2000critique}.
All statistical analyses are conducted using Python's \textit{SciPy}~\cite{virtanen2020scipy} and \textit{Scikit-posthocs}~\cite{terpilowski_scikit-posthocs_2019} libraries.

\begin{figure}[]
\centering

    \subfloat[]{
        \includegraphics[height=3.7cm]{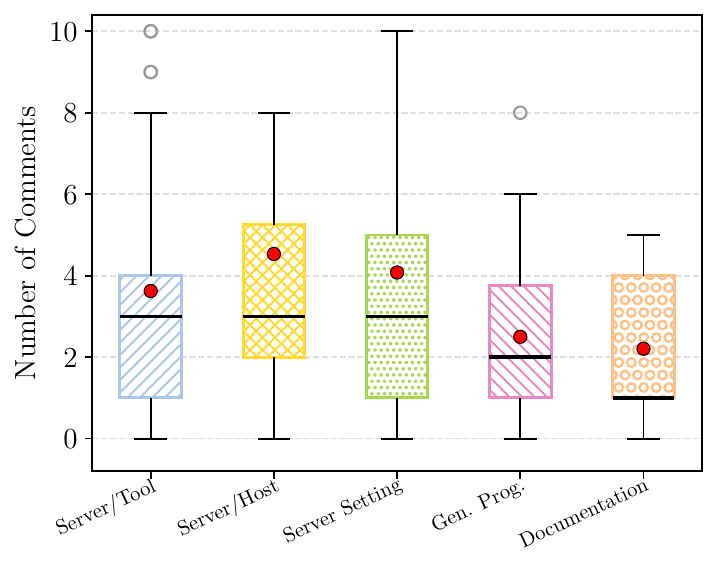}
        \label{fig:5g_metrics:sub2}
        }
    \subfloat[]{
        \includegraphics[height=3.7cm]{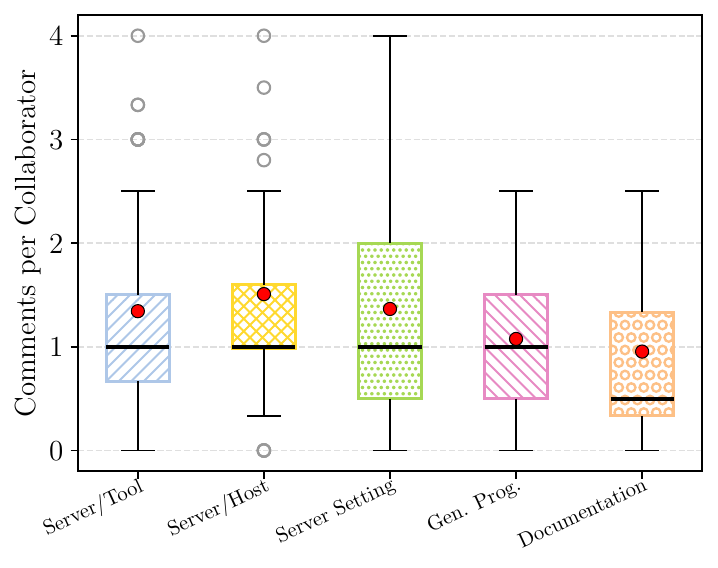}
        \label{fig:5g_metrics:sub4}
        }
    \subfloat[]{
        \includegraphics[height=3.7cm]{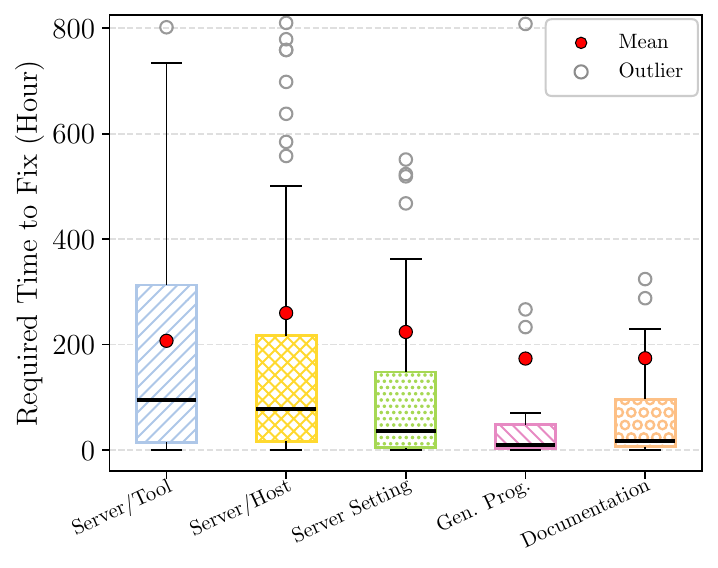}
        \label{fig:5g_metrics:sub1}
        }
        
    \subfloat[]{
        \includegraphics[height=3.7cm]{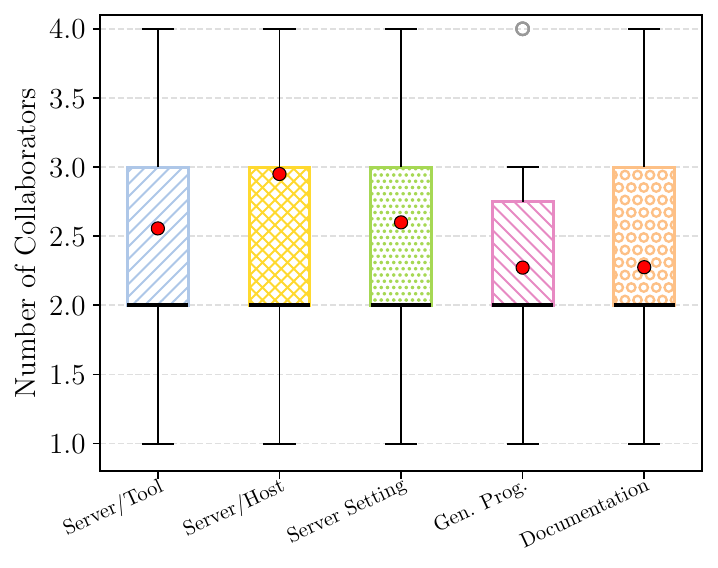}
        \label{fig:5g_metrics:sub3}
        }
    \subfloat[]{
        \includegraphics[height=3.7cm]{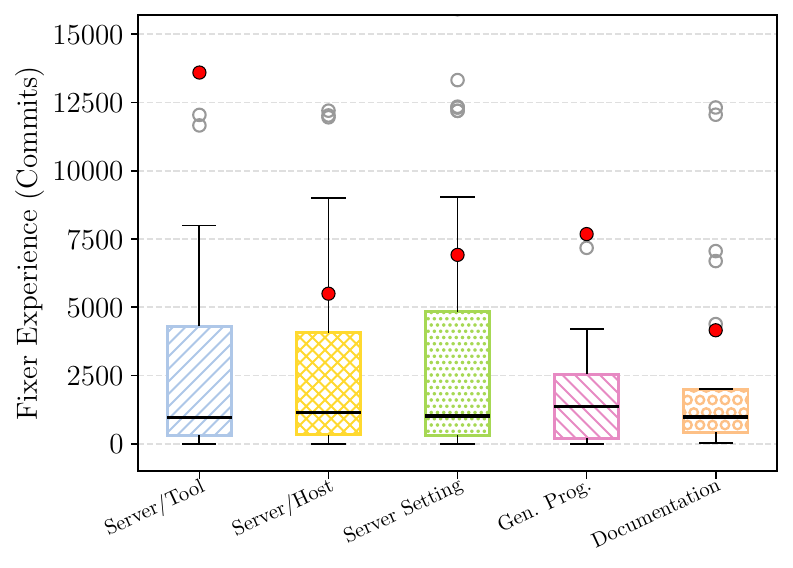}
        \label{fig:5g_metrics:sub5}
        }
     \caption{\centering Comparison of faults' characteristics for the 5 main categories of bugs in terms of metrics including (a) number of comments on the issue, (b) number of comments per collaborator, (c) required time to fix the fault, (d) number of collaborators on the issue, and (e) experience level of the developer who fixes the issue}
     \label{fig:5g_metrics}
\end{figure}

\subsubsection{Differences Across MCP Fault Categories}

Fig.~\ref{fig:5g_metrics} presents a comparison of these metrics across the various MCP-related fault categories identified in our taxonomy.
Table~\ref{tbl:Kruskal–Wallis} also reports the \textit{p}-values and corresponding effect sizes obtained from comparing the different metrics across categories of MCP-related faults.
As shown in the table, MCP-related fault types exhibit significant differences in the number of comments, the number of comments per collaborator, and the time required to fix an issue. The corresponding effect sizes are small, indicating that these results should be interpreted as systematic tendencies rather than large separations among fault categories. In other words, the categories differ most consistently in discussion intensity and resolution time, while no statistically significant differences are observed for the number of collaborators or fixer experience.
Table \ref{tbl:dunn} presents the results of the post-hoc pairwise comparisons conducted using \textit{Dunn’s} test, along with the corresponding effect sizes. It is important to note that the table reports only those pairs of fault types for which statistically significant differences are observed. Pairs that do not exhibit significant differences are omitted for clarity.

\begin{table}[]
\small
\caption{\centering Comparison of issue characteristics across five main types of MCP-related faults using the \textit{Kruskal–Wallis} test - (S = \textit{Small} effect size)}
    \centering
    \begin{tabular}{p{6cm} r r}
        \hline
        Metrics & \textit{p-Value} & \textit{effect size}    \\
        \hline
        Number of comments & \textbf{0.0114} & \textbf{0.0310 (S)}\\
        Number of comments/collaborator & \textbf{0.0273} & \textbf{0.0262 (S)}\\
        Required time to fix & \textbf{0.0008} & \textbf{0.0451 (S)} \\
        Number of collaborators & 0.2549 & 0.0128 \\
        Experience of fixer & 0.9987 & 0.0002 \\
        \hline

    \end{tabular}
        \label{tbl:Kruskal–Wallis}
        \vspace{-1em}
\end{table}

\begin{table}[b]
\small
\caption{\centering Post-hoc pairwise comparisons of pairs of fault types (\textit{Dunn} test) - (M) = Medium effect size}
    \centering
    \begin{tabular}{p{4.5cm} l l r r}
        \hline
        Metrics & Group \textit{a} & Group \textit{b}  &\textit{p-Value} & \textit{effect size}   \\
        \hline
        Number of comments & Documentation & Server/host config & 0.0136 & 0.3056 (M) \\
        Number of comments/collaborator & Documentation & Server/host config & 0.0137 & 0.2989 (M)\\
        Required time to fix & General programming & Server/host config & 0.0281 & 0.3023 (M)\\
        Required time to fix & General programming & Server/tool config & 0.0164 & 0.2939 (M)\\
        \hline
            
    \end{tabular}
        \label{tbl:dunn}
        \vspace{-1em}
\end{table}

According to the results of \textit{Dunn’s} post-hoc test, the number of comments and the number of comments per collaborator exhibit statistically significant differences between the \textit{Documentation} and \textit{Server/host configuration} fault categories. As illustrated in Fig.~\ref{fig:5g_metrics:sub2}, issues categorized as \textit{Server/host configuration} receive, on average, a higher number of comments than \textit{Documentation} issues. This difference suggests that server/host configuration faults require more diagnostic negotiation than documentation faults. In many server/host fault
issues, the failure is bound to the user's host environment, local configuration, or execution setting; maintainers may not be able to directly access or reproduce that environment and therefore need to request required logs, ask follow-up questions, compare host settings, or iterate over partial reproduction attempts. 
By contrast, documentation issues are usually more localized: once the missing or incorrect information is identified, the fix often requires less back-and-forth technical diagnosis. This interpretation is consistent with prior studies indicating that documentation issues in software repositories are generally less technically complex than functional bug-related issues, as they often concern formatting, clarity, or tool support rather than in-depth debugging or fault localization~\cite{aghajani2020software}. In contrast, bug reports frequently involve detailed failure descriptions, reproduction steps, and stack traces, which are associated with increased developer effort and longer resolution times~\cite{soltani2020significance}.
Similarly, Fig.~\ref{fig:5g_metrics:sub4} shows that the average number of comments per collaborator is higher for \textit{Server/host configuration} issues than for \textit{Documentation} issues. This pattern indicates that the extra discussion is not merely caused by more people joining the issue discussion,
but by more intensive back-and-forth among the participants already involved in diagnosing the fault. In other words, server/host configuration faults appear to increase the depth of diagnostic interaction rather than the breadth of participation.

In addition, \textit{Required time to fix} is identified as another metric exhibiting statistically significant differences across fault categories, as reported in Table~\ref{tbl:dunn}. As shown in Fig.~\ref{fig:5g_metrics:sub1}, issues classified under the \textit{General programming} category require the shortest average resolution time. This finding can be explained by the well-established nature of \textit{General programming} issues, for which developers possess extensive experience and mature debugging practices. In contrast, MCP-related faults are relatively novel, and the software engineering community has limited experience in diagnosing and resolving such issues.
Furthermore, issues in the \textit{Server/host configuration} and \textit{Server/tool configuration} categories tend to require more time to fix than \textit{General programming} issues, as they involve interactions among multiple software components (e.g., between the \textit{MCP server} and \textit{MCP host}, or between server-defined tools and the resources they invoke). Server/host configuration faults often require reconstructing how the server is launched, configured, and observed within a particular host-side environment. Server/tool configuration faults can similarly require checking whether tool schemas, tool execution behavior, and returned tool responses match what the host expects. Thus, the recurring challenge is not only that MCP faults span components, but also that part of the failure context may be outside the maintainer's directly observable environment. This observation aligns with previous empirical findings showing that \textit{General programming} issues are typically less complex and require fewer resources to resolve than AI-related faults~\cite{morovati2024bug}.

For the \textit{number of collaborators} and \textit{fixer experience}, the Kruskal--Wallis test does not identify statistically significant differences across the five categories. Although the boxplots suggest some descriptive variation, these metrics should not be interpreted as reliable differentiators among MCP fault categories. This result suggests that the main measurable differences among categories are not simply explained by more developers being involved or by more experienced developers fixing particular categories, but rather by the amount and intensity of discussion and the time needed to resolve the issue.

\subsubsection{MCP and Non-MCP Fault Comparison}

\begin{figure}[]
\centering
  
     \subfloat[]{
        \includegraphics[height=4cm]{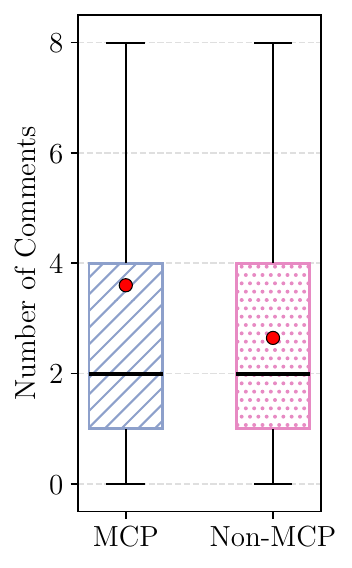}
        \label{fig:mcp-non-mcp:num-comments}
        }
        \hfill
    \subfloat[]{
        \includegraphics[height=4cm]{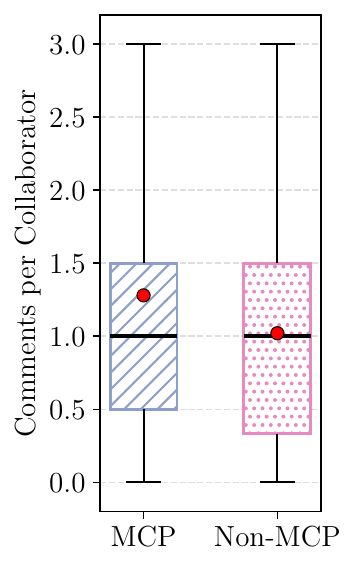}
        \label{fig:mcp-non-mcp:comment-collab}
        }
        \hfill
     \subfloat[]{
        \includegraphics[height=4cm]{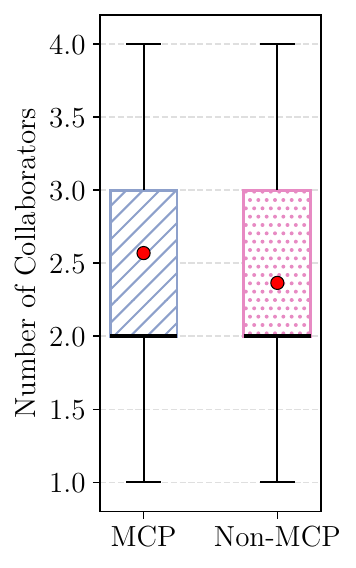}
        \label{fig:mcp-non-mcp:num-collab}
        }
        \hfill
    \subfloat[]{
        \includegraphics[height=4cm]{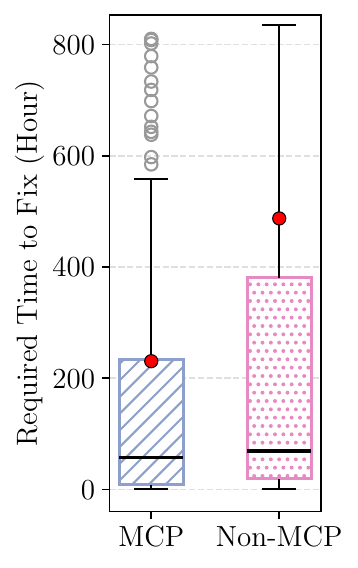}
        \label{fig:mcp-non-mcp:fix-time}
        }
        \hfill
    \subfloat[]{
        \includegraphics[height=4cm]{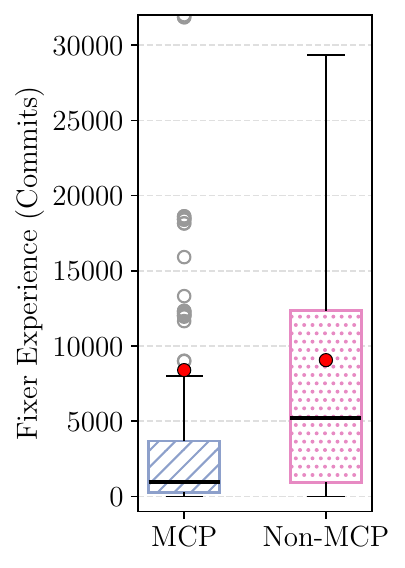}
        \label{fig:mcp-non-mcp:fix-exp}
        }
     \caption{Comparison of faults' characteristics for MCP vs. Non-MCP}
     \label{fig:mcp-non-mcp}
\end{figure}

To highlight the differences between MCP and non-MCP faults, we construct the comparison group from the same bug-related issue pool used in the preceding analysis. Specifically, after identifying 407 MCP-related issues from the 3,282 bug-related closed issues, the remaining 2,875 issues constitute the non-MCP comparison population. Because these issues come from the same repository corpus, collection period, and filtering pipeline, the comparison controls for project context while distinguishing MCP-related faults from other bug-related issues in MCP server repositories. We randomly sample 339 issues from this finite population using a 95\% confidence level, a 5\% confidence interval, and a conservative 50\% population proportion. Thus, the comparison group should be interpreted as representative of non-MCP faults within our collected MCP-server repository corpus, rather than as a general sample of all software faults. Fig.~\ref{fig:mcp-non-mcp} shows the comparison of the mentioned metrics for MCP and non-MCP faults.
To assess the significance of differences between MCP and non-MCP faults, we compare MCP and non-MCP faults using the \textit{Mann–Whitney U} test~\cite{mann_test_1947},
a non-parametric test that assesses whether two independent groups differ in their distributions. Since \textit{Vargha–Delaney A$_{12}$} is directly derived from the U statistic and remains fully interpretable for two-group nonparametric comparisons, we compute \textit{Vargha–Delaney A$_{12}$} effect size to quantify the magnitude of these differences. Table~\ref{tbl:mcp-non-mcp} highlights the \textit{p-value} and effect size of comparison between MCP and non-MCP faults, with respect to various metrics.

Regarding the \textit{number of comments} and the \textit{number of comments per collaborator}, Figures~\ref{fig:mcp-non-mcp:num-comments} and~\ref{fig:mcp-non-mcp:comment-collab} show that MCP-related faults exhibit higher average values for both metrics; Table~\ref{tbl:mcp-non-mcp} further shows that both differences are statistically significant with small effect sizes. At the same time, the number of collaborators does not differ significantly between MCP and non-MCP faults. Taken together, these results suggest that MCP-related faults require more intensive discussion, but not necessarily larger discussion groups. The additional effort appears as deeper back-and-forth among the involved developers, rather than simply as more developers joining the issue discussion.

This pattern is consistent with the nature of MCP faults observed during manual analysis. Many MCP-related issues cross boundaries between the server, host, tool, and user environment. When a fault depends on a user's local host configuration, launch command, authentication setup, transport mode, or logs, maintainers may need several rounds of clarification before they can reproduce or diagnose the problem. Similarly, when a server-defined tool depends on a local command, application, database, external API, or service, failures may arise from the invoked component, its returned payload, authentication or authorization state, network communication, version compatibility, or the server-side handling of the response before it is returned to the host. These cross-component dependencies help explain why MCP faults generate more discussion per issue and per participant.

In contrast, the \textit{required time to fix} differs significantly between MCP and non-MCP faults, but the effect size is negligible. Therefore, we interpret this result cautiously: MCP-related faults may receive faster attention because they affect core MCP functionality, but the practical magnitude of this difference is limited in our data. Thus, higher discussion intensity does not necessarily imply longer calendar resolution time; MCP-related issues may receive concentrated attention because they affect core server usability. The clearest MCP/non-MCP difference concerns fixer experience, where non-MCP faults are fixed by more experienced developers with a medium effect size. One plausible explanation is role specialization within MCP-server projects: developers with broader repository history may handle general-purpose implementation, infrastructure, or legacy code issues, whereas MCP-specific faults may be addressed by contributors focused on newer MCP-facing components.

In summary, RQ3 shows that MCP-related faults differ from non-MCP faults less by the number of people involved and more by the nature of the diagnostic discussion. MCP-related faults generate more discussion per issue and per participant, suggesting that developers must reason through protocol-specific interactions and component boundaries. Because most observed effects are small, these results should be used to identify recurring friction points rather than to rank fault categories by absolute difficulty. These findings suggest that MCP debugging support should focus on reducing diagnostic uncertainty, for example by improving error messages, configuration validation, host-server compatibility checks, and tool-interaction logging.

\begin{table}[]
\small
\caption{\centering Comparison of MCP-related and non-MCP-related faults across issue characteristics (\textit{Mann–Whitney U} test) - (S) = Small effect size, (N) = Negligible effect size}
    \centering
    \begin{tabular}{p{5cm} r r }
        \hline
        Metrics  &\textit{p-Value} & \textit{effect size}    \\
        \hline
        Number of comments & \textbf{0.0002} & \textbf{0.5746 (S)}\\
        Number of comments per collaborator & \textbf{0.0004} & \textbf{0.5719 (S)}\\
        Required time to fix & \textbf{0.0081} & \textbf{0.4454 (N)} \\
        Experience of fixer & \textbf{0.0000} & \textbf{0.3343 (M)} \\
        Number of collaborators & 0.2860 & 0.5202 \\
        \hline
            
    \end{tabular}
        \label{tbl:mcp-non-mcp}
        \vspace{-1em}
\end{table}

\begin{keyfindingsbox}{RQ3 - Key Findings}
\begin{itemize}
    \item Across the five MCP fault categories, significant differences are observed for the number of comments, comments per collaborator, and required time to fix, but the overall effect sizes are small.
    \item \textit{Server/host configuration} faults require more diagnostic discussion than \textit{Documentation} faults, especially when maintainers cannot directly reproduce the user's host-side environment.
    \item \textit{Server/host configuration} and \textit{Server/tool configuration} faults take longer to resolve than \textit{General programming} faults, reflecting cross-component debugging across hosts, servers, tools, and user environments.
    \item Compared with non-MCP faults, MCP-related faults involve more intensive discussion but not significantly more collaborators; the required-time-to-fix difference is practically negligible, while non-MCP faults are fixed by more experienced developers with a medium effect size.
\end{itemize}
\end{keyfindingsbox}

\subsection{Practical Implications and Actionable Insights}
\label{subsec:implications}

The three RQs 
jointly show that MCP reliability problems are not merely ordinary implementation defects in a new library. They often arise at protocol-mediated boundaries: between a server and its host, between server-defined tools and the local or remote resources they invoke, between returned tool payloads and host-side parsing, and between project-level configuration and user-specific execution environments. Thus, the practical implication is not simply to add more tests, but to make these boundary assumptions explicit, testable, and visible during debugging. The resulting implications concern MCP server developers, host developers, and developers of server-defined tools. We organize the practitioner-facing implications below as five actionable insights (A1--A5).

\paragraph{A1. Move boundary checks earlier in development.}
Because most observed faults are concentrated in \textit{Server/Tool Configuration}, \textit{Server/Host Configuration}, and \textit{Server Setting}, MCP server systems 
should treat host--server--tool boundaries as first-class test targets. In addition to unit-testing tool logic, developers should add startup and integration tests that exercise supported transport modes, host configuration templates, environment-variable handling, tool registration, tool-call execution, schema validation, and response parsing. Importantly, these checks should use the same launch path used by MCP hosts, rather than only invoking the server directly, because many failures are introduced by host-managed paths, arguments, environment variables, working directories, and transport definitions.

\paragraph{A2. Validate responses after successful tool invocation.}
The survey results indicate that \textit{tool response handling} is one of the most frequently encountered practitioner-facing faults. This suggests that successful tool registration and invocation should not be treated as sufficient evidence of tool reliability. 
MCP software systems should also validate response contracts, output-size limits, structured error objects, schema conformance, and host-side parsing of returned tool results. Since server-defined tools may wrap local commands, applications, databases, APIs, or other services, tests should include abnormal returned payloads from those invoked resources and verify that the server converts them into valid, diagnosable MCP responses. For developers of server-defined tools, this means documenting expected response shapes and failure modes; for server developers, it means checking behavior under malformed, empty, oversized, partial, or unexpected tool results.

\paragraph{A3. Reduce diagnostic back-and-forth through reproducible failure reports.}
Many MCP failures cannot be reproduced from the symptom alone because the relevant context may be split across the host, server launch configuration, SDK version, transport mode, local environment, and returned tool payload. This makes issue-report completeness especially important for maintainers. Accordingly, projects should use MCP-specific issue templates that collect the host name and version, server version, SDK version, transport mode, launch command, operating system, dependency manager, redacted configuration file, working directory assumptions, relevant environment variables, startup logs, failing tool schema, and a minimal example of the returned tool response. For host developers, the corresponding implication is to expose these details more clearly in error reports and debugging panels, so that users can attach a reproducibility package rather than reconstructing the environment through several rounds of clarification.

\paragraph{A4. Make protocol-boundary failures observable.}
Reproducible reports are still only one side of the debugging problem; MCP implementations should also make boundary failures observable at runtime. In particular, MCP server developers should strictly separate protocol messages from diagnostic logs, for example by ensuring that JSON-RPC messages are emitted only through the expected protocol channel and that logs are routed to \texttt{stderr} or a dedicated logging sink rather than \texttt{stdout}. Host developers should surface parsing, logging, transport, and protocol-channel failures as actionable boundary errors, including the failing transport mode, parse error, relevant log channel, and redacted malformed payload when available, rather than exposing them as generic connection or startup failures. This shifts part of the diagnostic burden from post hoc issue discussion to explicit runtime evidence about where the host--server--tool interaction failed.

\paragraph{A5. Prioritize using prevalence, severity, and repair effort together.}
The taxonomy and survey results show that the most common faults are not always the most severe or the most effort-intensive. \textit{Documentation} and \textit{tool response handling} are frequently encountered and should receive continuous documentation, examples, and regression-test attention. In contrast, 
according to the practitioner survey, \textit{host connection mismatch} has the highest severity index, while \textit{tool discovery \& registration} has the highest share of critical responses; these faults can block core MCP functionality and should therefore be included in release-gating checks. \textit{Tool authentication} has the highest effort index, indicating that authentication flows, authorization assumptions, credential configuration, and provider integration should receive design reviews and explicit compatibility tests before release. In short, MCP teams should prioritize faults using prevalence, severity, and effort together rather than treating issue frequency as the only signal.

\section{Related Works}
\label{sec:relatedwork}
This section presents and discusses the related literature.
\subsection{MCP and Agent Interoperability}
Although MCP has emerged as a de facto standardized interface for enabling LLMs to interact with external tools and resources in a structured manner, recent survey work situates it within a broader ecosystem of agent interoperability protocols—including Agent Communication Protocol (ACP), Agent-to-Agent (A2A), and Agent Network Protocol (ANP). This line of work highlights key architectural trade-offs across client–server, peer-to-peer, and decentralized designs~\cite{ehtesham_survey_2025}. Complementary reviews of A2A further examine aspects such as capability discovery, task-driven workflows, JSON-RPC-based communication, and deployment challenges in multi-agent environments~\cite{ray_review_2025}. Taken together, these studies frame MCP not as an isolated standard, but as a foundational layer for tool invocation and typed data exchange within increasingly modular and composable LLM-powered systems.

Beyond conceptual overviews, recent work has examined the structure, security, and evolution of the MCP ecosystem. Hou et al.~\cite{hou_model_2025} analyze the MCP lifecycle—including server creation, operation, and update phases—and identify associated security and privacy risks. Guo et al.~\cite{guo_measurement_2025} provide the first large-scale empirical measurement study of MCP marketplaces, revealing structural risks such as dependency monocultures, uneven maintenance, and inflated project listings. At the systems level, ScaleMCP proposes dynamic tool synchronization and retrieval mechanisms to improve agent autonomy and scalability in MCP-enabled environments~\cite{lumer_scalemcp_2026}. These works collectively advance understanding of MCP’s architecture, adoption trajectory, and system-level optimization.

Complementing architectural and ecosystem analyses, recent benchmarks evaluate how effectively LLM agents leverage MCP servers in realistic settings. MCP-Universe introduces execution-based evaluation across real-world MCP servers and highlights long-horizon reasoning and unknown-tool challenges~\cite{luo_mcp-universe_2025}. Similarly, MCPGAUGE systematically measures proactivity, compliance, effectiveness, and computational overhead in MCP-augmented LLM interactions~\cite{song_help_2025}. While these studies focus on interoperability design, ecosystem growth, and performance benchmarking, they largely emphasize capability and evaluation rather than reliability. In particular, there remains limited empirical investigation into recurring fault patterns in MCP servers and host integrations as observed in real-world development contexts. Our work addresses this gap by providing a systematic taxonomy of MCP-related faults derived from large-scale analysis of open-source issue reports.

\subsection{Failures in LLM-Based and Tool-Augmented Systems.}
As LLM-powered systems increasingly incorporate external tools and orchestration frameworks, researchers have begun to systematically study their failure modes. 
Several works propose taxonomies of failures in tool-augmented or agentic LLM systems. 
Winston and Just~\cite{winston_taxonomy_2025} introduce a taxonomy of failures in tool-augmented LLMs (TaLLMs), analyzing root causes in systems such as \textit{Gorilla} and \textit{Chameleon}. Islam et al.~\cite{islam_when_2026} conduct a large-scale empirical study of 1,187 bug reports from developer forums and repositories, identifying common bug types and structural components in LLM agents. 
Similarly, Xue et al.~\cite{xue_characterization_2025} characterize 1,026 bugs across mainstream LLM agent workflow frameworks, providing taxonomies of root causes and symptoms. 
At a broader systems-level perspective, Vinay~\cite{vinay_failure_2025} introduces a comprehensive taxonomy of latent failure modes in production-grade LLM applications, encompassing phenomena such as multi-step reasoning drift, erroneous tool invocation, and regressions induced by model version updates.
These studies collectively frame LLM reliability as an emerging software engineering challenge and highlight recurring issues in agent orchestration and tool integration.

Complementary work examines defects at other layers of the LLM-based software stack. Tian et al.~\cite{tian_taxonomy_2025} provide a taxonomy of prompt defects, analyzing how specification, formatting, and context errors degrade system behavior. At the infrastructure level, Liu et al.~\cite{liu_first_2026} present the first empirical study of bugs in LLM inference engines, identifying 28 root causes across 929 real-world issues, while Jiang et al.~\cite{jiang_foundation_2025} analyze bug characteristics and testing practices in widely used LLM libraries and frameworks,
revealing a shift toward interface-oriented and API-related defects. Survey work on agentic programming further synthesizes architectural patterns and challenges in tool-integrated LLM systems~\cite{wang_ai_2025}, reinforcing the need for systematic reliability engineering across the stack.

In parallel, a growing body of benchmarking research evaluates the robustness of tool-enabled LLM agents under realistic conditions. Benchmarks such as LiveMCP-101~\cite{yin_livemcp-101_2025}, MCP-AgentBench~\cite{guo_mcp-agentbench_2025}, and LiveMCPBench~\cite{mo_livemcpbench_2025} stress-test multi-tool orchestration in dynamic environments and report significant performance variability and failure rates. However, these efforts primarily assess task success and behavioral performance rather than mining and classifying real defect reports. Moreover, while prior studies characterize failures at the prompt, agent-framework, or infrastructure layers, limited attention has been given to protocol-mediated integration faults arising specifically from MCP server–host interactions. Our study complements existing taxonomies by isolating and empirically characterizing MCP-related faults grounded in large-scale analysis of open-source issue reports.

\section{Threats to Validity}
\label{sec:threat}
As with any empirical study, our work is subject to threats to validity. In this section, we discuss potential threats related to construct validity, internal validity, external validity, conclusion validity, and reliability, and describe the mitigation strategies we adopted to reduce their impact.

\subsection{Construct Validity} 
\label{subsec:threats_construct}
A primary threat to construct validity of this study arises from using GitHub closed issues as a proxy for real faults in MCP servers. Issue reports may include feature requests, usage questions, or environment-specific misconfigurations rather than genuine software faults. To mitigate this threat, we explicitly classify closed issues into multiple categories (bugs, feature requests, questions, and \textit{other}) and focus our analysis exclusively on issues labeled as bug-related (Section \ref{subsubsec:classification}). This practice is consistent with prior empirical software engineering studies~\cite{herzig2013s,long2025learning}.

Another potential threat to the construct validity stems from the use of LLM-assisted issue classification and summarization. 
Although 
LLMs may introduce misclassification errors or omit relevant contextual details,  
we evaluate several state-of-the-art LLMs against a manually labeled ground-truth dataset and select the best-performing model 
(Section \ref{subsubsec:classification}), 
to mitigate this risk.
Furthermore, LLMs are used only as preprocessing aids for classification and summarization; the identification of MCP-related issues and the construction of the fault taxonomy rely on subsequent clustering and manual inspection (Sections \ref{subsubsec:clustering} and \ref{subsubsec:manual_labeling}, respectively).
Finally, the fault characteristics used in our analysis (e.g., required time to fix, number of comments, number of collaborators, and experience of the person who fixes the fault) may not fully capture fault severity or complexity. To mitigate this threat, we adopted metric definitions that are widely used in prior empirical studies~\cite{kononenko2018studying,morovati2024bug} and interpreted results using both statistical significance and effect sizes, avoiding over-reliance on any single measure.

\subsection{Internal Validity}
\label{subsec:threats_internal}
The clustering process introduces a potential threat to the internal validity of this paper. Topic modeling and clustering techniques require design choices (such as the embedding model, dimensionality reduction method, and the number of clusters) that may affect the resulting groupings.
Furthermore, due to the infeasibility of manually inspecting all 3,282 bug-related issues, clustering was employed as a filtering mechanism to identify potentially MCP-related issues. This approach introduces the risk of false negatives, whereby some MCP-related issues may have been assigned to clusters primarily characterized by non-MCP topics and consequently excluded from manual labeling.
To mitigate these threats, we follow a principled clustering strategy by using \textit{HDBSCAN} to estimate a suitable number of clusters and then applying \textit{KMeans} within the \textit{BERTopic} framework (Section \ref{subsubsec:clustering}). 
To further reduce the likelihood of missing relevant issues, we adopt a conservative inclusion strategy by manually inspecting and including any issue whose summarized title or content explicitly referenced \textit{`MCP'}, even if it did not belong to an MCP-focused cluster. Although it is possible that a small number of MCP-related issues were not captured, the mitigation strategies we employed significantly reduce the risk that any such omissions meaningfully affect the resulting taxonomy. 
Manual annotation also poses a potential threat to internal validity due to researcher bias and subjectivity, particularly given the open coding approach adopted to construct the taxonomy.
To mitigate this threat, two independent raters with domain expertise labeled the MCP-related issues following established qualitative research guidelines (Section \ref{subsubsec:manual_labeling}). 
The raters worked independently and met regularly to discuss emerging labels, resolve disagreements, and refine the codebook. When consensus could not be reached, a third rater adjudicated the final decision. After finalizing the taxonomy, both raters re-examined all issues to ensure consistency with the agreed-upon definitions, reducing the likelihood of individual bias influencing the results.

\subsection{External Validity}\label{subsec:threats_external}
A primary source of threat to external validity arises from the scope of the analyzed MCP servers. Our study focuses exclusively on open-source MCP servers implemented using the Python MCP SDK. 
Since Python is currently the most widely used programming language for general-purpose development~\cite{python_popularity} as well as for implementing LLM-based software systems~\cite{twist2025study}, we believe that the findings of this study are applicable to a substantial portion of existing MCP server implementations.
Besides, although our taxonomy is derived from MCP server development, some identified fault categories (such as configuration, dependency management, and host–server interaction issues) may be relevant to other LLM-based tool-integration frameworks. However, we do not claim direct generalizability beyond MCP-based systems, and future work is needed to assess the applicability of the taxonomy to other LLM integration paradigms.

\subsection{Conclusion Validity} 
\label{subsec:threats_conclusion}
A potential threat to conclusion validity arises from the statistical analysis of fault characteristics, particularly given the non-normal distribution of the studied metrics (e.g., required time to fix, number of comments, and number of collaborators). To mitigate this threat, we explicitly test for normality and relied on non-parametric statistical tests that do not assume normal distributions, including the \textit{Kruskal--Wallis} test for multi-group comparisons and the \textit{Mann--Whitney} U test for pairwise comparisons (Section~\ref{subsec:fault_characteristics}). These tests are widely recommended in empirical software engineering research, particularly when the assumptions of normality are not satisfied.

Another threat concerns the risk of over-interpreting statistically significant results, especially in the presence of large sample sizes, where small differences may become statistically significant without being practically meaningful. To address this issue, we consistently report effect sizes alongside \textit{p-value}s, including \textit{$\eta^2$} for omnibus tests and \textit{Vargha--Delaney $A_{12}$} for pairwise comparisons (Section~\ref{subsec:fault_characteristics}). This allows us to assess the magnitude and practical relevance of observed differences rather than relying solely on statistical significance.
The use of multiple statistical comparisons introduces an additional threat by increasing the likelihood of Type~I errors. We mitigate this risk by applying \textit{Bonferroni} correction during post-hoc analyses (Section~\ref{subsec:fault_characteristics}), following established best practices in empirical research.
Finally, our study is observational in nature, and therefore does not support causal claims about the relationship between fault types and their observed characteristics. We mitigate this threat by framing our findings as descriptive and comparative trends rather than causal relationships, and by avoiding claims that extend beyond what the statistical analyses can support.

\subsection{Reliability and Reproducibility} \label{subsec:threats_reliability}

A potential threat to reliability arises from the use of external tools and non-deterministic components in the analysis pipeline, including LLMs used for issue classification and summarization, embedding models used for topic modeling, and clustering algorithms with stochastic behavior (Sections~\ref{subsubsec:classification} and~\ref{subsubsec:clustering}). Variations in model versions, API updates, or random initialization may lead to slight differences in intermediate results if the study is repeated. To mitigate this threat, we report the specific models, versions, configurations, and prompt templates used throughout the study, and we apply consistent settings across all experiments. In addition, while LLMs and clustering are used to assist preprocessing, the final identification of MCP-related issues and the construction of the fault taxonomy rely on manual inspection and consensus-based labeling, which reduces sensitivity to minor variations in automated outputs.

Another threat concerns the replicability of the data collection and analysis process. To address this, we make all datasets, and analysis artifacts used in this study publicly available through a replication package~\cite{replication_package}. 
Although complete determinism cannot be guaranteed due to the inherent stochasticity of some components, the transparency of our methodology, the availability of all artifacts, and the use of well-established tools substantially reduce threats to reliability and reproducibility.

\section{Conclusion and Future Works}
\label{sec:conclusion}

This paper presents the first large-scale empirical study of real-world faults in Model Context Protocol (MCP) software systems. To achieve this, we manually analyzed and labeled 407 MCP-specific closed GitHub issues collected from 279 MCP server repositories. The resulting analysis led to the construction of a taxonomy of MCP-specific faults comprising five high-level categories, with faults related to \textit{server setting}, \textit{server/tool configuration}, and \textit{server/host configuration} emerging as the most prevalent in practice. To assess the validity of the proposed taxonomy, we conducted a survey with 41 MCP practitioners, whose responses confirm both the completeness and practical relevance of the identified MCP-specific fault categories. Furthermore, our empirical analysis highlights clear differences between the characteristics of MCP-specific faults and those of non-MCP faults in MCP-based software systems.
The findings of this study provide actionable guidance for practitioners by identifying the most error-prone and critical components of MCP servers, enabling developers to prioritize robustness and reliability during system design and implementation. From a research perspective, this work establishes a foundational understanding of MCP-related faults and can serve as a baseline for empirical studies, benchmarks, and automated techniques for fault detection, diagnosis, prioritization, and repair in MCP-based systems.
As future work, we plan to expand the scope of this study by analyzing MCP clients and incorporating longitudinal data to examine how fault patterns evolve as MCP matures. Moreover, we intend to investigate benchmark construction, boundary-contract conformance, and learning-based approaches for proactively identifying and mitigating MCP-related faults, thereby further improving the reliability and trustworthiness of AI-enabled software systems built upon MCP.

\bibliographystyle{ACM-Reference-Format}
\bibliography{_bibliography}

\clearpage
\appendix

\section{Representative Taxonomy Examples}
\label{app:taxonomy_examples}
Table~\ref{tbl:taxonomy_examples} summarizes the representative issue examples that were originally discussed within the taxonomy description. We moved these examples to the appendix to keep the main Results section focused on the taxonomy structure and cross-category interpretation while preserving traceability to concrete issue evidence.

\begingroup
\scriptsize
\setlength{\tabcolsep}{3pt}
\renewcommand{\arraystretch}{1.15}
\begin{longtable}{@{}L{0.17\textwidth} L{0.48\textwidth} L{0.29\textwidth}@{}}
\caption{Representative examples illustrating the taxonomy categories.}
\label{tbl:taxonomy_examples}\\
\toprule
\textbf{Fault type} & \textbf{Representative issue illustration} & \textbf{What the example illustrates} \\
\midrule
\endfirsthead
\toprule
\textbf{Fault type} & \textbf{Representative issue illustration} & \textbf{What the example illustrates} \\
\midrule
\endhead
\midrule
\multicolumn{3}{r}{\textit{Continued on next page}}\\
\endfoot
\bottomrule
\endlastfoot
\textit{Server Dependency} &
A dependency conflict prevented installation of \texttt{dicom-mcp}; the report noted that \textit{``pynetdicom has a conflicting dependencies''}, producing a \texttt{ResolutionImpossible} error~\cite{ex_server_dependency}. &
Server startup and installation can fail before MCP execution begins when required package versions cannot be resolved. \\
\textit{Server Platform Incompatibility} &
On Windows, an MCP memory service was \textit{``treating log messages as JSON''} and produced \textit{``multiple parsing errors''}, a behavior linked to platform-specific stream buffering~\cite{ex_server_platform}. &
The same server behavior can fail differently across operating systems because protocol streams and process I/O are platform-sensitive. \\
\textit{Server Deployment} &
A pre-built Docker image failed at startup because a required package was missing; the issue argued that the \textit{``package should be included as a dependency''} in the image~\cite{ex_server_deployment}. &
Packaging and deployment artifacts must contain the same dependencies expected by the server runtime. \\
\textit{Server Access Synchronization} &
Concurrent tasks caused \textit{``asyncio event loop contamination''} and occasional \textit{``unrecoverable deadlock''} conditions, motivating process isolation~\cite{ex-server-resource-sync}. &
Shared runtime resources can require isolation when multiple operations execute concurrently. \\
\textit{Server Resource Handling} &
Stale project-state metadata caused initialization to fail with \textit{``Project name 'XXXXX' already exists''} because the name appeared to point to another repository~\cite{ex_server_resource_handling}. &
Server resource state can become inconsistent with the user's current workspace. \\
\textit{Server Logging} &
A lowercase log-level value triggered startup validation failure; the error stated that \texttt{log\_level} should be \textit{``DEBUG, INFO, WARNING, ERROR or CRITICAL''}~\cite{ex_server_logging}. &
Logging configuration can become part of the server's startup contract. \\
\textit{Server Network Configuration} &
Local callback servers launched for authentication were \textit{``never getting shut down''}, leaving \textit{``multiple http servers listening''} after repeated attempts~\cite{ex_server_network_config}. &
Auxiliary network components need explicit lifecycle management. \\
\textit{API Usage} &
In \texttt{aws-api-mcp-server}, the server initialized a HuggingFace/SentenceTransformer model-loading API with a restrictive \texttt{local\_files\_only} setting, preventing the required model from being downloaded when it was absent or outdated in the local cache~\cite{ex_server_api_usage}. &
Server-side API parameters and model-loading configuration must match the expected execution environment, especially when MCP servers rely on external libraries or service APIs during initialization. \\
\textit{Server Infrastructure} &
A memory-saving operation became extremely slow; the user wrote that it \textit{``took a century''}, and maintainers later attributed the behavior to an infrastructure issue~\cite{ex_server_infrastructure}. &
Some observed MCP failures originate from external infrastructure rather than server code. \\
\textit{Tool Call/Execution} &
Tool invocations failed with \textit{``Invalid type for parameter `params' in tool''} because tools exposed Pydantic \texttt{BaseModel} wrappers that clients \textit{``couldn't parse''}~\cite{ex-tool-callexecution}. &
Tool parameter schemas and wrapper types must match the argument structure expected by MCP clients during invocation. \\
\textit{Tool Registration} &
A \texttt{version} tool appeared twice because it was registered through a new class and through a leftover path that \textit{``manually registers''} the same tool~\cite{ex_tool_registration}. &
Registration logic can expose duplicate or stale tools when migration paths are incomplete. \\
\textit{Tool Exposure/Discovery} &
Implemented tools decorated with \texttt{@mcp.tool()} produced \textit{``Tool not found''} errors and did \textit{``not appear in the available tools list''} because capabilities were not exposed~\cite{ex_tool_exposure}. &
Tool implementation alone is insufficient; discovery metadata must also be correct. \\
\textit{Tool Response Handling} &
The server returned full embedding vectors in tool responses; each fact included \textit{``1536 float values''}, inflating responses from roughly 5K to more than 250K tokens~\cite{ex_tool_response}. &
Post-invocation response content can create severe cost, performance, and usability problems even when the tool call succeeds. \\
\textit{Tool Authorization} &
A knowledge-base query returned a \textit{``403 Error''} because default reranking required additional Bedrock permissions~\cite{ex_tool_authorization}. &
Tool behavior may require permissions beyond those needed for direct or simpler API calls. \\
\textit{Tool Authentication} &
OAuth failed when fields documented as \textit{``optional''} were left as placeholder credentials; real credentials were required even for read-only operations~\cite{ex_tool_authentication}. &
Authentication documentation and implementation assumptions can diverge. \\
\textit{Tool Dependency} &
A configured tool \textit{``doesn't show up in the app''} until startup logs revealed a runtime error; upgrading to \texttt{Python~3.13} resolved the incompatibility~\cite{ex_tool_dependency}. &
Tool-specific runtime requirements can prevent tools from loading even when the server is configured. \\
\textit{Tool Connection Configuration} &
Concurrent Gmail batch requests overloaded the SSL layer, producing repeated \textit{``[SSL] record layer failure''} and \texttt{TypeError: terminated} errors~\cite{ex_tool_connection}. &
Connections used by server-defined tools can fail under concurrent or high-load interaction patterns. \\
\textit{Host Connection Synchronization} &
A Docker-based MCP server logged \textit{``RuntimeError: Received request before initialization was complete''} when a request arrived too early~\cite{ex_host_conn_sync}. &
Hosts must wait for server initialization before issuing tool calls. \\
\textit{Host Connection Configuration Mismatch} &
A user attempted to connect using SSE, but the maintainer clarified that \textit{``the MCP server doesn't support SSE''} and advised selecting \textit{STDIO}~\cite{ex_host_conn_mismatch}. &
Host and server transport assumptions must match. \\
\textit{Incorrect Host Connection Setting} &
Clients were \textit{``trying to connect to the wrong /messages''} endpoint instead of the required \texttt{/mcp/messages} route~\cite{ex_host_conn_incorrect}. &
Small endpoint-format differences can break host--server handshakes. \\
\textit{Session Handling} &
Multiple Claude Code sessions against the same server started \textit{``looking at a different repository''}, indicating that session data was mixed across workspaces~\cite{ex_host_conn_session}. &
Concurrent host sessions require explicit state isolation. \\
\textit{Host's Server Configuration} &
Claude Desktop could not launch the server, raising \texttt{spawn wikipedia-mcp ENOENT}, because its \textit{``limited PATH environment''} did not include the directory where \texttt{pip install} placed the executable; specifying the full executable path in the server configuration fixed the problem~\cite{ex_host_server_config}. &
Host server configurations must account for the host application's launch environment, including executable paths that may differ from the user's shell. \\
\textit{LLM} &
An LLM \textit{``corrected''} an entity identifier after one successful Home Assistant command, causing later calls to target a nonexistent entity~\cite{ex_host_llm}. &
LLM-mediated tool use can fail when the model rewrites structured identifiers. \\
\textit{Host Dependency} &
A host update caused all tool schemas to fail with \textit{``this schema is not valid''} because it no longer supported a union of object types~\cite{ex_host_dependency}. &
Host dependency or schema-support changes can break otherwise unchanged MCP tools. \\
\textit{Host Logging} &
An MCP server \textit{``does prints in console''}, making it \textit{``unusable with the STDIO transport''} until logs were routed to \texttt{stderr}~\cite{ex_host_logging}. &
Diagnostic output must be separated from protocol messages. \\
\textit{Host's Hook Configuration} &
Hook configuration contained \textit{``hardcoded paths from the original developer's machine''}, breaking lifecycle automation for new users~\cite{ex_host_hook}. &
Host lifecycle hooks must avoid machine-specific configuration. \\
\textit{Host Authorization} &
A deployed app \textit{``goes to 'crashed' state''} because the service principal lacked required catalog-level privileges such as \texttt{USE CATALOG}~\cite{ex_host_authorization}. &
Host-mediated operations may require platform permissions beyond the immediate server configuration. \\
\textit{Documentation} &
Sample configuration used \texttt{REDIS\_PSW} instead of \texttt{REDIS\_PWD}, causing \textit{``authentication errors with the sample config''}~\cite{ex_documentation}. &
Documentation examples can propagate incorrect configuration into user environments. \\
\textit{General Programming} &
An MCP server \textit{``crashed on startup due to a syntax error''} because nested double quotes in an f-string caused a parsing error~\cite{ex_general_programming}. &
Some faults are ordinary implementation mistakes rather than MCP-specific integration faults. \\
\end{longtable}
\endgroup

\clearpage
\section{Used Prompt Templates}
\label{app:prompt}

\lstdefinelanguage{PromptText}{
  morecomment=[l]{\#}
}

\lstdefinestyle{promptstyle}{
  language=Python,
  basicstyle=\ttfamily\scriptsize,
  keywordstyle=\color{kw}\bfseries,
  stringstyle=\color{str},
  commentstyle=\color{gray!70},
  showstringspaces=false,
  breaklines=true,
  columns=fullflexible,
  keepspaces=true,
  frame=single,
  framerule=0.5pt,
  rulecolor=\color{frame},
  backgroundcolor=\color{bg},
  xleftmargin=6pt,
  xrightmargin=6pt,
  captionpos=b,
  escapeinside={(*@}{@*)}
}

\subsection{Issue Classification}
The following templates list our 4-group and binary classification prompts respectively, which were used with all LLMs described in section \ref{subsubsec:classification}. \texttt{\textcolor{kw}{\{Texts\}}} are replaced with actual contents as context knowledge for each issue.
\begin{lstlisting}[style=promptstyle,caption={Prompt template for 4-group issue classification },label={lst:prompt-classification1}]
{"role": "system", "content": """You are an expert GitHub issue triage assistant. Your task is to classify GitHub issues into one of the following categories:

(A) Bug - A problem or error in the code causing crashes, incorrect behavior, or failures.
(B) Feature Request - A request to add new functionality or improve existing features.
(C) Question - A request for help, clarification, or information about usage or behavior.
(D) Other - Anything that does not clearly fit into the categories above.

You must return only the best label from this list in the format ((X)). Do not explain your reasoning.

### Example 1:
Title: App crashes on file save
Description: When saving, the app crashes with a segmentation fault. This started after the 1.4 update.
Comments: I'm getting this too after the update!  Seems related to the new auto-save logic.

Label: ((A))

### Example 2:
Title: Add dark mode option
Description: It would be great to have a dark theme for working at night.
Comments: Yes, please! Eye strain is a real issue.  Hope this gets prioritized.

Label: ((B))

### Example 3:
Title: How do I enable autosave?
Description: I'm not sure how to enable autosave in the settings.
Comments: It's only in the Pro version.  Docs are a bit unclear.

Label: ((C))


### Example 4:
Title: Just wanted to say thank you
Description: This tool has been amazing. Looking forward to what's next!
Comments: Appreciate the kind words!

Label: ((D))

### Now classify this issue: 

"""},

{"role": "user", "content": """\
Title: (*@\kw{\{title\}}@*)

Description: (*@\kw{\{body\}}@*)

Comments: (*@\kw{\{comments\}}@*)

Label: "}
\end{lstlisting}

\clearpage

\begin{lstlisting}[style=promptstyle,caption={Prompt template for binary issue classification},label={lst:prompt-classification2}]
{"role": "system", "content": """You are an expert GitHub issue triage assistant. Your task is to classify GitHub issues into one of the following two categories:

(A) Bug - A problem or error in the code causing crashes, incorrect behavior, or failures.
(B) Irrelevant - An issue that is invalid, stale, duplicated, not actionable or not reproducible.

You must return only the best label from this list in the format ((X)). Do not explain your reasoning.

### Example 1:
Title: App crashes on file save
Description: When saving, the app crashes with a segmentation fault. This started after the 1.4 update.
Comments: I'm getting this too after the update!  Seems related to the new auto-save logic.

Label: ((A))

### Example 2:
Title: Crash only happened once
Description: I experienced a crash when opening a file, but now I can't reproduce it.
Comments: Might be a local issue.  Works fine for me too.

Label: ((B))

### Now classify this issue: """},

{"role": "user", "content": """\
Title: (*@\kw{\{title\}}@*)

Description: (*@\kw{\{body\}}@*)

Comments: (*@\kw{\{comments\}}@*)

Label: "}
\end{lstlisting}

\subsection{Issue Summarization}

The following template lists our summarization prompt used with all LLMs described in section \ref{subsubsec:classification}. \texttt{\textcolor{kw}{\{Texts\}}} are replaced with actual contents as context knowledge for each issue.

\begin{lstlisting}[style=promptstyle,caption={Prompt template for issue summarization},label={lst:prompt-summarization}]
{"role": "system", "content": 
"""You are a helpful software engineering assistant.
Your job is to analyze closed GitHub issues and determine what the issue was really about.
Many issue titles are vague or misleading, so always read the body and comments carefully."""},

{"role": "user", "content":  
"""\
Title: (*@\kw{\{title\}}@*)

Body: (*@\kw{\{body\}}@*)

Comments: (*@\kw{\{comments\}}@*)

Task:
1. Provide a concise SUBJECT label that summarizes the actual issue, like a bug category or specific technical error (e.g., 'NullPointerException in ConfigLoader', 'Docker build fails on Alpine').
2. Write a 1-2 sentence SUMMARY explaining what the issue was and what caused or resolved it.

Respond in the following format:
Subject: <subject here>
Summary: <summary here>
"""}
\end{lstlisting}

\end{document}